\documentclass[preprint]{JASAnew}

\begin{document}



\title{WHIS: Hearing impairment simulator based on the gammachirp auditory filterbank }
\author{Toshio Irino}
\email{irino@wakayama-u.ac.jp}
\affiliation{Faculty of Systems Engineering,  Wakayama University, 930 Sakaedani Wakayama, 640-8510, Japan}


\begin{abstract}
\nolinenumbers 
\setlength{\baselineskip}{18pt}

A new version of a hearing impairment simulator (WHIS) was implemented based on a revised version of the gammachirp filterbank (GCFB),  which incorporates fast frame-based processing, absolute threshold (AT), an audiogram of a hearing-impaired (HI) listener, and a parameter to control the cochlear input-output (IO) function. The parameter referred to as the compression health $\alpha$ controlled the slope of the IO function to range from normal hearing (NH) listeners to HI listeners, without largely changing the total hearing loss (HL). 
The new WHIS was designed provide an NH listener the same EPs as those of a target HI listener.The analysis part of WHIS was almost the same as that of the revised GCFB, except that the IO function was used instead of the gain function. We proposed two synthesis methods: a direct time-varying filter for perceptually small distortion and a filterbank analysis-synthesis for further HI simulations including temporal smearing. We evaluated the WHIS family and a Cambridge version of the HL simulator (CamHLS) in terms of differences in the IO function and spectral distance. The IO functions were simulated fairly well at $\alpha$ less than 0.5 but not at $\alpha$ equal to 1. Thus, it is difficult to simulate the HL when the IO function is sufficiently healthy. This is a fundamental limit of any existing HL simulator as well as WHIS. The new WHIS yielded a smaller spectral distortion than CamHLS and was fairly compatible with the previous version.

\end{abstract}


\maketitle

\nolinenumbers 
\setlength{\baselineskip}{18pt}

\vspace{-2pt}
\section{Introduction}
\label{sec:Introduction}
\vspace{-2pt}

As many countries approach the super-aging society status, the number of hearing-impaired (HI) listeners may increase. It is crucial to develop next-generation assistive devices that can compensate for the difficulties faced by individual HI listeners. For this purpose, it is essential to effectively specify the dysfunctions without a heavy experimental load. Many psychoacoustic experiments have been conducted to clarify the dysfunctions using relatively simple stimulus sounds, such as sinusoids and noise \,\citep{moore2013introduction}. In addition, many speech sound experiments have been performed, although they have mainly been restricted to intelligibility tests, such as speech-in-noise tests.
However, in experiments conducted with elderly HI listeners, it is not easy to specify whether the deterioration factor is located on the periphery, the auditory pathway, or cognition. This is because of the huge variability among HI listeners in terms of both audiograms and cognitive factors.

To resolve this problem, at least partially,
a hearing loss (HL) simulator was developed to specify the effects of the peripheral dysfunction, such as elevation of absolute threshold (AT) and loudness recruitment, on speech intelligibility \,\citep{villchur1974simulation}.
Normal-hearing (NH) listeners could evaluate the speech intelligibility of the HL-simulated sounds, which might correspond to what HI listeners perceive. In the HL simulator, input signals were decomposed into frequency bands by using a set of linear filters, and then expanded and synthesized to obtain the simulated signal.
\citet{moore1993simulation} initially introduced auditory filters into an HL simulator for loudness recruitment and a more precise simulation of the frequency selectivity. The auditory filters were psychoacoustically estimated by conducting notched-noise (NN) masking experiments\,\citep{patterson1976auditory} for both NH and HI listeners\,\citep{patterson1982deterioration,glasberg1986auditory}. An rounded exponential (roex) filter was used in the estimation.
Spectrum smearing was introduced to evaluate the effect of the bandwidth widening in an HI listener's auditory filter \,\citep{baer1993effects,baer1994effects}. In addition, a unified version was developed to include loudness recruitment and spectrum smearing \,\citep{nejime1997simulation}. This Cambridge version of the HL simulator is referred to as CamHLS. CamHLS was used in a study of the upper limit of temporal delay in hearing aids \,\citep{stone1999tolerable}. Recently, CamHLS has also been used in the base line system of ``Clarity Prediction Challenge'' (CPC1) \,\citep{claritychallenge}, which is a competition conducted to develop a new objective measure for hearing-aid signal processing. 
There are other HL simulators as well. For example, HeLPS v2 \,\citep{zurek2007hearing} is commercially available and includes the simulation of loudness recruitment. 
We also developed another type of HL simulator \,\citep{irino2013accurate,nagae2014hearing, irino2020gammachirp}, which is referred to as the Wakayama-University Hearing Impairment Simulator (WHIS),
which is based on a dynamic compressive gammachirp filterbank (GCFB)\,\citep{irino2006dynamic}.
The gammachirp filter in GCFB is a time-domain filter, unlike the roex filter, which accounts for the NN masking thresholds fairly well\,\citep{irino1997time, patterson2003extending}. The gammachirp requires smaller number of coefficients than the roex filter\,\citep{unoki2006comparison}. Based on this psychoacoustical backgrounds, GCFB was also used in an HL simulator functionally similar to CamHLS \,\citep{hu2011simulation} and a real-time HL simulator \,\citep{grimault2018real}.
WHIS was designed to control the degree of the compression in the cochlear input-output (IO) function rather than to simulate the loudness recruitment directly. WHIS has been used in various experimental studies as described in the Discussion section. 

However, HL simulators have been mostly used in speech intelligibility studies. This is probably because the simulated sounds contain more or less distortion components, which preclude precise psychoacoustic experiments and sound quality evaluations.
Although the distortion is unavoidable due to nonlinear signal processing in nature, it can be reduced to a sufficiently small level with sophisticated processing.
For this purpose, a goodness measure of the HL simulator is required. It is also important to know the fundamental limit of the HL simulator. To the best of our knowledge, there have been no reports on such measures and comparisons between HL simulators. 
Although these HL simulators were developed to simulate the peripheral dysfunction, more central temporal resolution or temporal modulation transfer function (TMTF) \,\citep{bacon1985temporal} is an important factor for speech perception\,\citep{drullman1994effect}. Thus, it is desirable to simulate both the peripheral and central dysfunctions within a unified framework.

In this study, a new version of WHIS, based on an improved version of GCFB, is developed to address these issues. The new GCFB is first explained as it improved in processing speed and incorporated the audiogram and compression characteristics of HI listeners. 
Then, the analysis and synthesis methods of WHIS are described based on the signal processing of GCFB.
WHIS and CamHLS are evaluated using spectral distance and IO function. Finally, we discussed the fundamental limit of HL simulators, estimation of the active and passive HLs, and applications of WHIS.

\section{Improvement in GCFB}
\label{sec:GCFB}
WHIS was developed based on a compressive gammachirp filter (cGC) \,\citep{irino2001compressive} and a dynamic compressive gammachirp filterbank (GCFB) \,\citep{irino2006dynamic}. 
For the implementation of the new WHIS, GCFB should be improved to meet the following WHIS specifications, 1) fast frame-based processing for an interactive user interface, 2) clear definition of the cochlear output level relative to the AT, and 3) incorporation of the audiograms and cochlear IO functions of HI listeners.

In this section, we first define a cGC filter then explain its improvement in detail. Although it is somewhat lengthy, it is essential to understand the concept of the new GCFB-based WHIS.

\subsection{cGC filter}
\label{sec:cGC}
The background of a cGC developed from the original gammachirp was reviewed by \citet{irino2020gammachirp}. 
The absolute frequency response of a cGC
\,\citep{irino2001compressive}, $|G_{CC}(f)|$, can be formulated as 

\begin{eqnarray}
|G_{CC}(f)|  &=& |G_{CP}(f)| \cdot H_{HPAF}(f) \label{eq:GCC=GCP_HPAF}.
\end{eqnarray}
Here
\begin{eqnarray}
|G_{CP}(f)| &=& a_{\Gamma}\,  |G_T(f)| \, \exp(c_1 \theta_1 ), 
\label{eq:GCP=aGTexp}\\
H_{HPAF}(f)  &=&  \exp(c_2 \theta_2 ),
    \label{eq:HPAF=expc2} \\
  \theta_1 &=& \arctan \biggl(\frac{f-f_{r_1}}{b_1 \rm ERB_N\it(f_{r_1})} \biggr), 
   \label{eq:Theta1} \\
    \theta_2 &=& \arctan \biggl(\frac{f-f_{r_2}}{b_2 \rm ERB_N\it(f_{r_2})} \biggr).
   \label{eq:Theta2} 
\end{eqnarray}
$|G_{CC}(f)|$ is a product of a passive gammachirp (pGC), $|G_{CP}(f)|$, and a high-pass asymmetric filter (HP-AF), $H_{HPAF}(f)$, which enables the level-dependent control of bandwidth and gain and is formulated as $\exp(c_1 \theta_1 )$.
$|G_{CP}(f)|$ is a product of a gammatone $|G_T(f)|$ and $\exp(c_1 \theta_1 )$ which introduces a frequency glide or chirp. 
The scalar value $a_{\Gamma}$ is the amplitude; $b_1$ and $b_2$ are bandwidth factors; $c_1$ and $c_2$ are chirp factors; and  $f_{r_1}$ and $f_{r_2}$ are the asymptotic frequency of pGC and the center frequency of HP-AF, respectively. $\rm{ERB_N}\it(f)$ is an equivalent rectangular bandwidth of NH listeners at frequency $f$ \,\citep{moore2013introduction}.

When the peak frequency of a pGC is $f_{p1}$ and the sound pressure level at the pGC output is estimated as $P_{gcp}$ on a dB scale, the center frequency of HP-AF,  the center frequency of HP-AF,  $f_{r2}$, is associated with $f_{p1}$ to introduce the level dependency of a cGC.
\begin{eqnarray}
    f_{r2} & = & f_{rat}(P_{gcp})\cdot f_{p1} \label{eq:fr2=fratPgcp}, \\
    f_{rat}(P_{gcp}) & = & f_{rat}^{(0)} + f_{rat}^{(1)}\cdot P_{gcp},
   \label{eq:fratPgcp=frat0} 
\end{eqnarray}
where $f_{rat}^{(0)}$ and  $f_{rat}^{(1)}$ are coefficients which are estimated together with the other parameters, $b_1$,$c_1$,$b_2$,and $c_2$, by conducting the NN masking experiments \,\citep{irino2001compressive}. The parameter values reported by \,\citep{patterson2003extending} are used in this study: $b_1=1.81$; $c_1=2.96$; $b_2=2.17$; $c_2=2.20$; $f_{rat}^{(0)}=0.466$; $f_{rat}^{(1)}=0.0109$.

\subsection{Introduction of frame-based processing into GCFB}
\label{sec:FrameBaseGCFB}
The original version of GCFB (hereafter, ${\rm GCFB_{v21}}$) was developed using the cGC formula to simulate cochlear filtering \,\citep{irino2006dynamic}. 
The level-dependent filtering in ${\rm GCFB_{v21}}$ required heavy computational costs because the filter coefficients in many channels were updated and convoluted with the input signal at each sample point. 
The duration required for this sample-by-sample processing was several tens to a hundred times of the input signal duration. Therefore, it could not be used as a background processor for the interactive human interface necessary in WHIS. This problem did not occur in the previous version of WHIS (hereafter ${\rm WHIS_{v22}}$), as it used several approximations.
In the new version of WHIS (hereafter ${\rm WHIS_{v30}}$), hearing loss (HL) was simulated on the basis of excitation patterns (EPs) from input signals, as described in the next section. 
The EPs can be calculated as short-time averaged levels of filterbank outputs and does not require sample-by-sample processing. Therefore, frame-based processing was introduced into the new version of GCFB (hereafter ${\rm GCFB_{v23}}$). As the filtering was performed in every frame of a few millisecond, it was possible to improve the processing speed. 
Although the temporal fine structure (TFS) was discarded in the frame-based processing, it could be calculated in the sample-by-sample circuit inherited from ${\rm GCFB_{v21}}$ if necessary.

\begin{figure*}[t]
\centerline{\includegraphics[width=1\linewidth,bb=0 0 963 611] {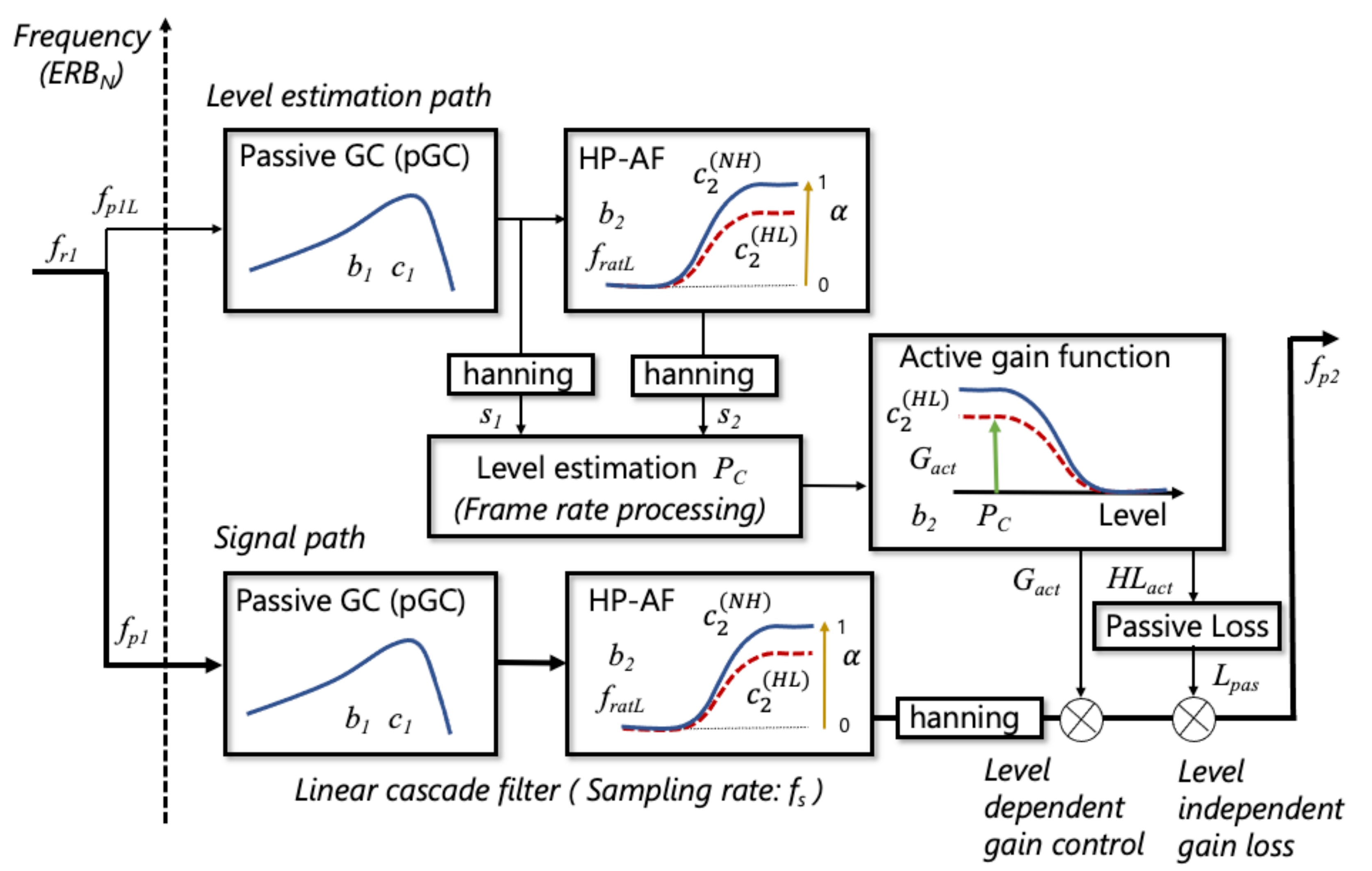}}
\caption{Block diagram of one channel of the frame-based GCFB, ${\rm GCFB_{v23}}$}
\label{fig:Diagram1ch_GCFBv23} 
 \vspace{-10pt}
 \end{figure*}

\subsection{Black diagram of GCFB}
\label{sec:BlockDiagramGCFB}

Figure \ref{fig:Diagram1ch_GCFBv23} shows a block diagram for one ${\rm GCFB_{v23}}$ channel. 
We provide an overview of the signal flow here before describing the individual components in the following sections. %
As in ${\rm GCFB_{v21}}$, there are two paths for level estimation (upper block) and signal flow (bottom block),
both of which have the same linear pGC and HP-AF filters, except for the center frequencies. 
Outputs of the level estimation path were used to estimate the signal level, $P_c$, by using hanning windows, where the rms level was calculated with a window length of 1\,ms and frame-shift of 0.5\,ms. The method for the level estimation using two signal sources (i.e., $s_1$ and $s_2$) was almost the same as that used for ${\rm GCFB_{v21}}$. 
The frame-based level, $P_c$, determined the gain value of an active gain function (right-middle block). The output of the linear filters in the signal path (bottom block) can also be summarized using the same hanning window.
The frame-based signal level was, then, controlled by the active gain function to produce the output.
The elapsed time for this frame-based processing was roughly the same as the input signal duration when using MATLAB. The short processing time is essential for WHIS and can be useful for various applications that do not require TFS.

\subsection{Introduction of the HL into GCFB}
\label{sec:HL2GCFB}
We explain the basic concept of how to introduce the HL into the model.

\subsubsection{Audiogram and HL}

 The peripheral HL of an HI listener can be modelled using the dysfunctions in the active and passive processes. We assumed the total HL, $HL_{total}$, as the sum of the active HL,  $HL_{act}$, and the passive HL, $HL_{pas}$, on a dB scale:
 \begin{equation}
   HL_{total} = HL_{act}+HL_{pas}.
   \vspace{-7pt}
    \label{eq:HLtotal_ACT+PAS}
\end{equation}
  \citet{moore1997model} proposed a similar equation on a dB scale, $ HL_{total} = HL_{OHC}+HL_{IHC}$,  where $HL_{OHC}$ is the HL caused by the outer hair cell (OHC) and $HL_{IHC}$ is that caused by the inner hair cell (IHC). Although the main concept for both equations is almost the same, Eq. \ref{eq:HLtotal_ACT+PAS} is preferred in this paper because the active process is not solely functioned by the OHC and the passive loss is not solely caused by the IHC dysfunction.

\subsubsection{Introduction of compression health}
\label{sec:CompressionHealth}

cGC in Eq.\ref{eq:GCC=GCP_HPAF} comprises pGC, which represents a passive filter, and HP-AF, which represents an active mechanism. We assumed that pGC is common to both NH and HI listeners because it is a broadband filter that could simulate the response of cochlear traveling wave at high sound pressure levels (SPLs). 
The dysfunction of the active process could be modelled by reducing the dynamic range of HP-AF in Eq.\ref{eq:HPAF=expc2},
which is determined by the coefficient $c_2$. The original value of $c_2$ 
is estimated from the NN thresholds of NH listeners ($c_2^{(NH)}$).
The $c_2$ values for HI listeners might be smaller because of the active HL ($c_2^{(HL)}$).
We introduced a coefficient $\alpha \; \{\alpha | 0 \le \alpha \le 1\}$ and defined the relationship between $c_2^{(NH)}$ and $c_2^{(HL)}$ as
\begin{eqnarray}
   c_2^{(HL)} & = & \alpha \cdot c_2^{(NH)}.
    \label{eq:c2HL=alpha}
\end{eqnarray}
At $\alpha = 1$, there is no dysfunction in the active process and this is the case for NH listeners. At $\alpha = 0$, the active function is completely damaged.
The value for an individual HI listener can be somewhat in the middle and frequency-dependent. We can control the $\alpha$ value based on the measurement or assumption about the degree of compression in the cochlear IO function, as described in the Discussion section.
Thus, the parameter $\alpha$ is referred to as ``compression health.''  Although the definition and value are different from those of the compression health $\alpha$ defined in ${\rm WHIS_{v22}}$\,\citep{irino2020gammachirp}, both $\alpha$s are highly correlated.

The HP-AF of an HI listener could be represented as
\begin{eqnarray}
  H_{HPAF}^{(HL)}(f)  &=&
  \exp(c_2^{(HL)} \theta_2 ) = 
    \exp(\alpha \cdot c_2^{(NH)} \theta_2 ).
   \label{eq:HHPAFHL=expC2HL} 
\end{eqnarray}
Two HP-AF curves when using $c_2^{(NH)}$ and $c_2^{(HL)}$ are depicted in the HP-AF blocks of Fig.\ref{fig:Diagram1ch_GCFBv23}. The dynamic range of HP-AF reduces as $\alpha$ reduces.
The amplitude spectrum of cGC for an HI listener is represented as
\begin{eqnarray}
  |G_{CC}^{(HL)}(f)| &=& |G_{CP}(f)|\cdot H_{HPAF}^{(HL)}(f). \label{eq:GCPHHPAFHL} 
\end{eqnarray}
As indicated in the previous section,
pGC and HP-AF are implemented as linear filters for speed-up. 
The frequency relationship between pGC and HP-AF is set by Eq.\ref{eq:fratPgcp=frat0} with $P_{gcp}$ fixed at approximately 50\,dB.
The peak gain is normalized to 0\,dB independently of the parameter $c_2^{(HL)}$. 
The impulse response is formulated as 
\begin{eqnarray}   
  g_{cc}^{(HL)}(t) & = & g_{cp}(t)*h_{HPAF}^{(HL)}(t),
    \label{eq:gcctgcpt}  
\end{eqnarray}
where $h_{HPAF}^{(HL)}(t)$ is an approximation filter of $H_{HPAF}^{(HL)}(f)$ which does not have phase information \,\citep{irino2006dynamic}. 
This linear cascade filter does not represent level-dependent gain and bandwidth in accordance with the input sound level.  However, the bandwidth difference between NH and HI listeners can be modeled by using $\alpha$.
The level-dependent gain is introduced by the active gain function (right-middle block), as shown in Fig.\ref{fig:Diagram1ch_GCFBv23}. 
The gain value $G^{(HL)}$ is determined from the estimated signal level $P_c(\tau)$, which is derived at each frame time $\tau$. Eq.\ref{eq:gcctgcpt} can be rewritten by using $G^{(HL)}(P_c(\tau))$ as
\begin{eqnarray}
  g_{cc}^{(HL)}(\tau)
  & = & W_{han}\{g_{cp}(t)*h_{HPAF}^{(HL)}(t)\}\cdot G^{(HL)}(P_c(\tau)),
  \label{eq:gcctgcptGain}
\end{eqnarray}
where $W_{han}\{.\}$ denotes the rms calculation performed using the hanning window, which resamples the signal sampling rate to the frame rate, as shown in Fig.\,\ref{fig:Diagram1ch_GCFBv23}.

The gain $G^{(HL)}(P_c(\tau)) $ is calculated by using Eqs. \ref{eq:HPAF=expc2} and \ref{eq:Theta2}, as follows:
\begin{eqnarray}
 G^{(HL)}(P_c(\tau)) & = & H_{HPAF}^{(HL)}(f_{p1},P_c(\tau))\\
    & = &   \exp\biggl\{c_2^{(HL)} 
	\cdot \arctan \biggl(\frac{f_{p_1}-f_{r_2}}{b_2 \rm ERB_N\it(f_{r_2})}
	\biggr)\biggr\} \nonumber\\
   f_{r_2} &=& f_{rat}(P_c(\tau))\cdot f_{p_1}
   \label{eq:fr2=fratPc}
\end{eqnarray}
where $f_{p1}$ is the peak frequency, $f_{r2}$ is the center frequency of HP-AF, and 
$f_{rat}(P_c(\tau))$ is similar to Eq.  \ref{eq:fr2=fratPgcp} but with the frame-based estimated level $P_c(\tau)$. 

The gain of the active process decreases as the input SPL increases, as observed in the cochlear IO function. 
When a high SPL that yields the active gain of 0\,dB is denoted as $P_{gain0}$ (e.g., 100\,dB), the active gain $G_{act}$ can be approximated as
\begin{eqnarray}
   G_{act}(P_c(\tau)) &=& \frac{H_{HPAF}(f_{p1},P_c(\tau))}{H_{HPAF}(f_{p1},P_{gain0})}.
   \label{eq:Gact}
\end{eqnarray}
This equation is valid for both NH and HI listeners because $P_{gain0}$ is assumed to be the same for both. The only difference is whether $c_2^{(NH)}$ or $c_2^{(HL)}$ should be used for calculating $H_{HPAF}(f_{p1},P_c(\tau))$.

\begin{figure}[t]
  \centerline{\includegraphics[width=0.8\linewidth]{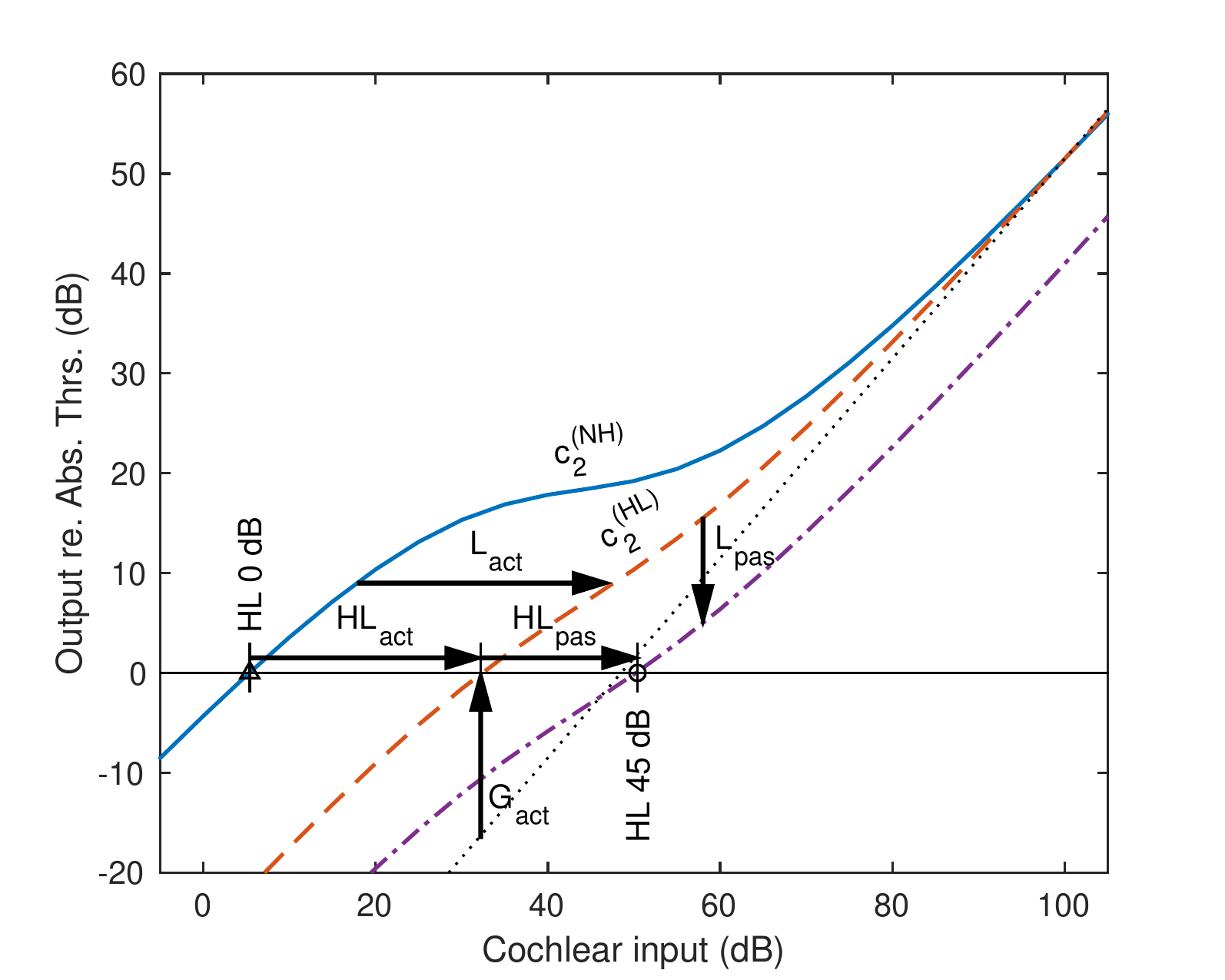}}
 \vspace{-5pt}
  \caption{Schematic plot of the cochlear IO function. The abscissa is the SPL (dB) at the cochlear input. The ordinate is the output level (dB) relative to the AT. The label ``HL\,0\,dB'' represents the input level corresponding to the AT of the average NH listener as defined in \,\citep{ansi2018specification}. The blue solid line represents the IO function of the NH listener. The black dotted line shows a linear relationship or 1:1. The orange dashed line shows the IO function when using $c_2^{(HL)}$ and $\alpha=0.5$ without any passive HL. Then, the threshold increases for $HL_{act}$ from HL\,0\,dB. When the passive HL exists, the line moves downward for $L_{pas}$. The purple dashed-and-dotted line shows the IO function of an HI listener whose hearing level is 45\,dB as an example. The AT further decreases for $HL_{pas}$. Thus $HL_{act}$+$HL_{pas}$ = 45\,dB in this case. $L_{act}$ is used in WHIS and is described by Eq.\,\ref{eq:LactFIO}.
  }

\label{fig:IOfunc_schematic}
\end{figure}

\subsubsection{Cochlear IO function and HL}
\label{sec:IOfunc_HL}
Figure \ref{fig:IOfunc_schematic} shows a schematic graph of the cochlear IO functions to explain the effects of the active and passive dysfunctions. The horizontal axis is the input level (SPL dB) to the cochlea, and the vertical axis is the output level (dB) relative to 
that for the AT or ``HL 0\,dB,'' which is calculated from the hearing level defined in \citet{ansi2018specification} and a transfer function between the middle ear and the cochlear input, which is used in a loudness model proposed by \citet{glasberg2006prediction}. 
``HL 0\,dB'' is located at the intersection of the IO function of the average NH listener (blue solid line labeled with $c_2^{(NH)}$) and the horizontal 0-dB line.
This IO function has linear growth in the low SPLs becomes compressive in the middle SPLs, and then converges onto a 1:1 linear function (dotted line) in the high SPLs. 
The hearing level of an HI listener is located on the right of ``HL 0\,dB'' on the horizontal 0-dB line.

First, let us consider a case in which the HL is solely caused by the active dysfunction. When we use $c_2^{(HL)}$ with $\alpha$ value of 0.5 instead of $c_2^{(NH)}$, in Eqs.\,\ref{eq:HHPAFHL=expC2HL} and \ref{eq:Gact}, the IO function becomes the red dashed line, which is steeper and less compressive.
The difference between the IO functions of $c_2^{(NH)}$ and $c_2^{(HL)}$ result in the elevation of the AT for $HL_{act}$ from ``HL\,0\,dB'' on the horizontal 0-dB line.
Then, by introducing the passive dysfunction, the IO function moves downward for $L_{pas}$ to the purple dashed-and-dotted line, which represents the case of, for example, an HI listener whose hearing level is 45\,dB (``HL\,45\,dB''). Then, the AT elevated for $HL_{pas}$. It is a rationale of Eq.\,\ref{eq:HLtotal_ACT+PAS} in which $HL_{total}$ (45\,dB in this case) is a sum of $HL_{act}$ and $HL_{pas}$.

Using this framework, the ratio between $HL_{act}$ and $HL_{pas}$ can be controlled without changing $HL_{total}$. We can specify the HL of an HI listener, $HL_{total}$, from the audiogram and $\alpha$ in advance. The active gain $G_{act}$ and $HL_{act}$ are automatically determined from the IO function as described above. Then, $HL_{pas}$ is determined as $HL_{pas}$ = $HL_{total} -  HL_{act}$ from Eq.\ref{eq:HLtotal_ACT+PAS}

\subsubsection{Implementation in GCFB}
\label{sec:IOfunc_Implementation}

In ${\rm GCFB_{v23}}$, the active gain, $G_{act}$, and the passive loss, $L_{pas}$ ($>0$), are calculated in the active gain function block and the successive block shown on the right side of Fig.\ref{fig:Diagram1ch_GCFBv23}.
The total gain applied to the output of the linear pGC and HP-AF filters is 
\begin{eqnarray}
   G_{total}(P_c(\tau))  & = &  G_{act}(P_c(\tau)) - L_{pas}
   \label{eq:Gtotal_Pc=Gact}
\end{eqnarray}
on a dB scale. Note that $G_{act}(P_c)$ is level-dependent while $L_{pas}$ is a constant that is determined from $HL_{pas}$ and the IO function.
This process is performed in every filterbank channel $n_{ch}$ with the estimated signal level $P_c(n_{ch}, \tau)$.
Equation\,\ref{eq:Gtotal_Pc=Gact} is rewritten more specifically as
\begin{eqnarray}
   G_{total}(n_{ch},P_c(n_{ch},\tau))  & = &  G_{act}(n_{ch},P_c(n_{ch},\tau)) - L_{pas}(n_{ch}).
     \label{eq:Gtotal_nch_Pc=Gact}
\end{eqnarray}

\begin{figure}[t]
  \centerline{\includegraphics[width=0.7\linewidth]{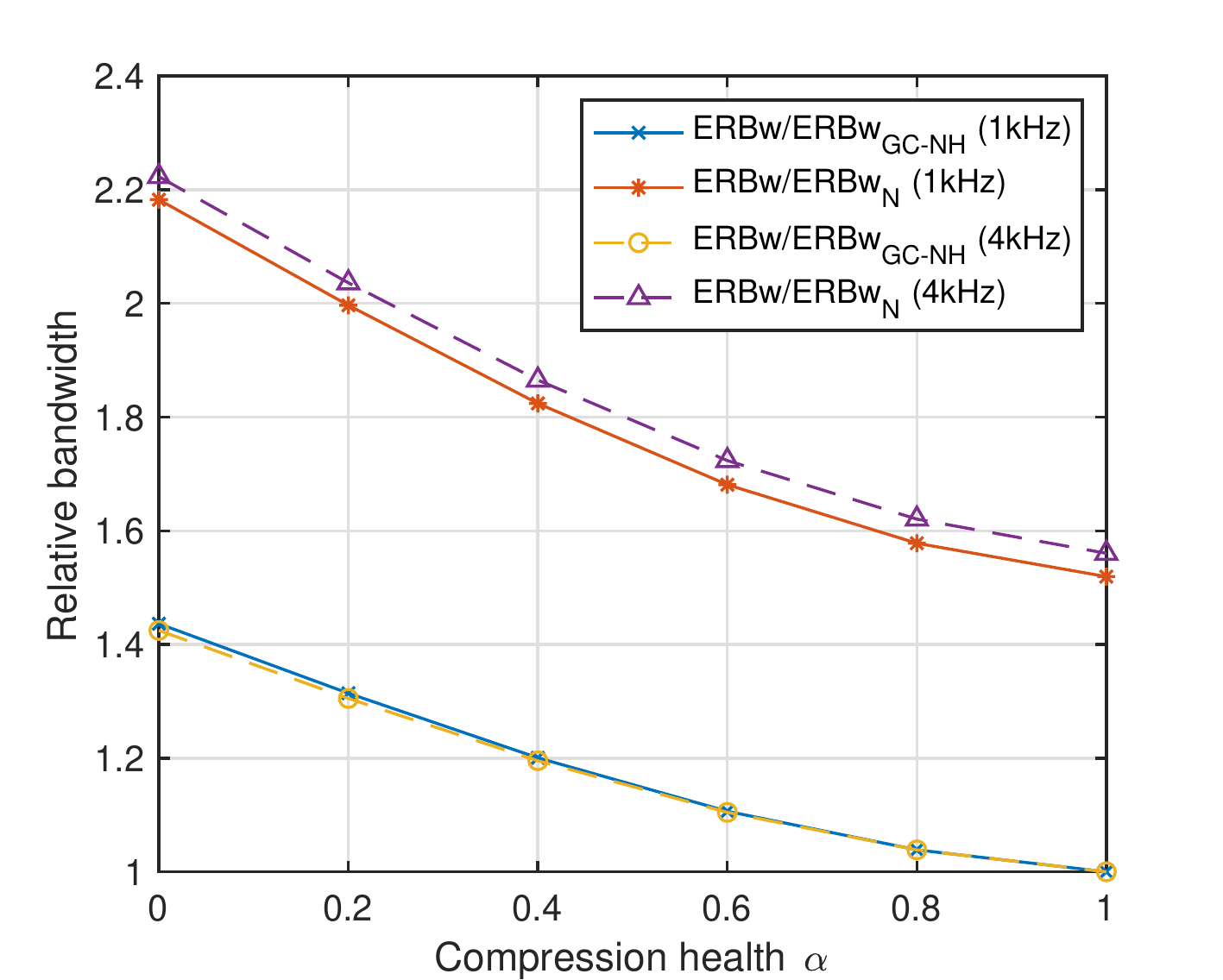}}
  \caption{Filter bandwidth as a function of the compression health, $\alpha$. Solid lines: 1\,kHz. Dashed lines: 4\,kHz. The two lower lines show the bandwidth relative to the case when $\alpha=1$ in ${\rm GCFB_{v23}}$. The two upper lines show the bandwidth relative to the standard $\rm ERB_N$ bandwidth of the average NH listener \,\citep{moore2013introduction}.}
\label{fig:Bandwidth_alpha}
\end{figure}

\subsection{Simulation results}
\label{sec:ResultGCFB}

The simulations using ${\rm GCFB_{v23}}$ are performed to evaluate the algorithms described above.

\subsubsection{Evaluation of bandwidth}
\label{sec:GCFB_Bandwidth}
As described previously,
the pGC and HP-AF filters are implemented as linear filters for speed-up. The filter shapes of HI and NH listeners are determined from the value of $\alpha$. Although the level dependence is not implemented in this formulation, it might not be entirely unreasonable, at least when simulating the HI listener's filter, because the nonlinearity is smaller in HI listeners than in NH listeners.

Figure \ref{fig:Bandwidth_alpha} shows the bandwidth of the cascade filter (pGC + HP-AF) as a function of $\alpha$ at signal frequencies of 1\,kHz and 4\,kHz.
The vertical axis is the relative bandwidth normalized by the bandwidth at $\alpha = 1$ (i.e., a completely healthy condition).
The lower curves show that, as $\alpha$ decreases from 1.0 to 0, the bandwidth increases gradually from 1.0 to 1.4 times. At $\alpha=0$, it is the bandwidth of the pGC filter because the frequency response of the HP-AF filter is unity (0\,dB), as indicated in Eq.\ref{eq:HHPAFHL=expC2HL}.

When the bandwidth is normalized by the standard $\rm{ERB_N}$ bandwidth\,\citep{moore2013introduction},
the relative bandwidth is 1.6 times wider than the value calculated above. 
This is because the bandwidth of the cGC filter estimated with the NN masking paradigm by \,\citet{patterson2003extending} is wider than $\rm{ERB_N}\it(f)$ at any SPL, and the current cascade filter is fixed at the characteristics of approximately 50\,dB SPL.
The bandwidth difference between NH and HI listeners is introduced in ${\rm GCFB_{v23}}$ although the degree of the difference is arguable. The values can be  compensated by changing the coefficients of cGC and GCFB, based on a more precise estimation of the auditory filer.

\begin{figure*}[t]
  \centerline{\includegraphics[width=1\linewidth]{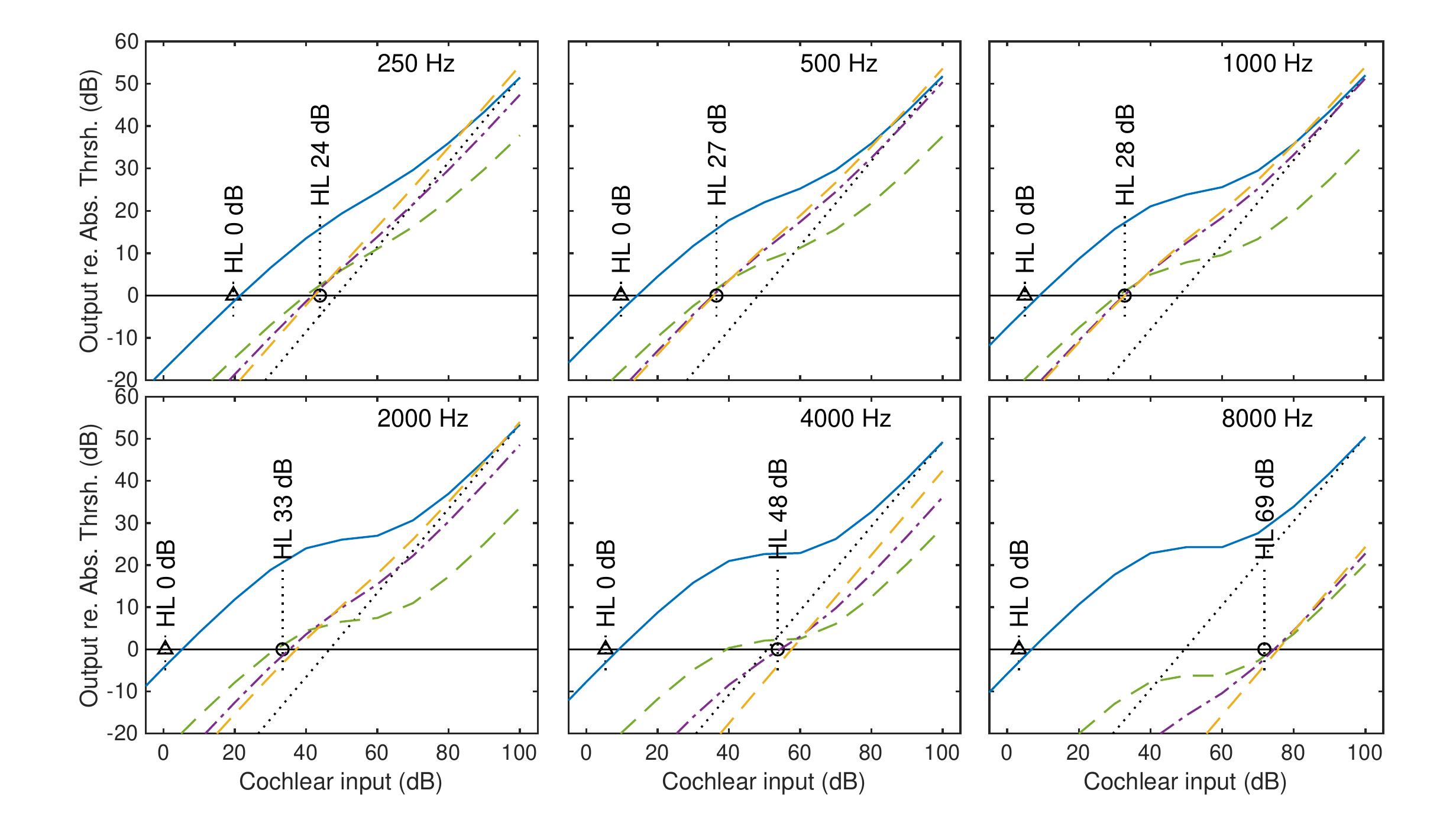}}
\caption{Simulation results on the IO function for frequencies between 250 and 8000\,Hz.
The abscissa is the SPL (dB) at the cochlear input. The ordinate is the output level (dB) relative to the AT. 
HL\,0\,dB corresponds to the AT of average NH listeners \,\citep{ansi2018specification}. The other HL values are derived from the average HL of 80-year-old male listeners reported by \,\citet{tsuiki2002nihon}.   Blue solid line: Average NH. Green dashed line: HL with the compression health $\alpha$ of 1.0. Purple dashed-and-dotted line: HL with $\alpha$ of 0.5. Orange dashed line: HL with $\alpha$ of 0.0. Black dotted line: The linear relationship (1:1).
}
\label{fig:IOfunc_ResultGCFB}
\end{figure*}

\subsubsection{Evaluation of IO function}
\label{sec:GCFB_IOfunction}

The IO function of ${\rm GCFB_{v23}}$ might not correspond to the schematic function shown in Fig.\ref{fig:IOfunc_schematic}, because of the introduction of several approximations, such as frame-based processing and the impulse response in Eq.\ref{eq:gcctgcpt}.
Therefore, we calculate the IO functions of ${\rm GCFB_{v23}}$ under the NH condition and three HI conditions at the audiogram frequencies between 125 and 8000\,Hz. The HL of the HI listener is set to the average value of 80 year-old males \,\citep{tsuiki2002nihon}. The value of $\alpha$ is set to 0.0, 0.5, and 1.0 to show the difference, because the audiogram does not provide  information about it. The practical values of $\alpha$ are different from 0, even when setting $\alpha=0$ in advance, because the minimum value of $\alpha$ is restricted by $HL_{total}$. The values are listed in Table\,\ref{tab:DiffHL_IO} in the parentheses (compensated $\alpha$).

Figure \ref{fig:IOfunc_ResultGCFB} shows the IO functions between 250 and 8000\,Hz. The input signals are sinusoids of 200\,ms duration with the level of every 10\,dB step from -10 to 100\,dB. The outputs are the maximum values of the EPs calculated by GCFB.

The IO functions of the NH listener (blue solid line) intersect the horizontal 0-dB line near the points labeled with ``HL 0\,dB'' (triangles). The differences between the zero cross points and ``HL 0\,dB'' are listed in the second row of Table \ref{tab:DiffHL_IO}. The maximum difference is 4.6\,dB at 2000\,Hz, and thus, smaller than 5\,dB, which is the resolution of a normal audiometry test.


\begin{table*}[t]
 \caption{Difference (dB) between the zero cross point of the IO function and the average hearing levels of NH and HI listeners in Fig.\ref{fig:IOfunc_ResultGCFB}. The hearing level of average 80-year-old HI listeners \,\citep{tsuiki2002nihon} is shown in the bottom row. The simulations were performed for $\alpha$ values of 0.0, 0.5, and 1.0. }
 \label{tab:DiffHL_IO}
 \centering
 \footnotesize
  \begin{tabular}{|c||c|c|c|c|c|c|c|c|}
   \hline
    Freq. (Hz) & 125 & 250 & 500 & 1000 & 2000 & 4000 & 8000\\
   \hline \hline
    NH ($\alpha=1$) & -2.1 & 1.3 & 4.2 & 3.9 & 4.6  & 3.6 & 3.5 \\
    \hline
    HI ($\alpha=1$) & -4.5 & -4.2 & -2.7  & -2.0 & -2.0 & -14.5 & 2.4 \\
    \hline
    HI ($\alpha=0.5$) & -5.4 & -2.2 & 1.1 & -0.2 & 1.9 & 0.5 & 2.6 \\
    \hline
    HI ($\alpha=0$) & -8.7 & -1.6 & -0.8 & -0.0 & 3.8 & 3.8 & 3.8 \\
    (compensated $\alpha$)  & (0.0) &  (0.03)  &  (0.37)  &  (0.43)  &  (0.33)  &  (0.0) &  (0.0) \\
    \hline
    Hearing level of HI & 23.5& 24.3 & 26.8 & 27.9 & 32.9 & 48.3 &  68.5\\
    \hline
  \end{tabular}
\end{table*}

The IO functions of the HI listener at $\alpha=1$  (green dashed line)  are shifted down for $L_{pas}$ without changing the shapes of those of the NH listener.
Therefore, they have compressive regions. The differences between the zero cross points and the HL values of the HI listener are listed in the third row of Table \ref{tab:DiffHL_IO}. The absolute differences are less than 5\,dB, except at 4000\,Hz, where the compressive region is very close to the horizontal 0-dB line, which increases the difference.
The IO functions are much steeper at $\alpha=0.5$  (purple dashed and dotted lines) and $\alpha=0.0$ (orange dashed lines) than at $\alpha=1.0$. The steep function can cause the loudness recruitment. The differences between the zero cross points and the HL values are listed in the fourth and fifth rows. Most of them are less than 5\,dB. The maximum absolute difference at 125\,Hz is less than 10\,dB.
The results demonstrated that the output level at the AT is set to 0\,dB in ${\rm GCFB_{v23}}$ fairly well for both NH and HI listeners.
It is confirmed that the AT is reasonably simulated independently of the $\alpha$ value, which determines the ratio of $HL_{act}$ and $HL_{pas}$.


\section{New implementation of WHIS}
\label{sec:WHIS}
A new version of WHIS (${\rm WHIS_{v30}}$) is developed based on the algorithm of ${\rm GCFB_{v23}}$ described in the previous section.

\subsection{Objective of the HL simulator}
\label{sec:ObjectiveHLS}
The first question is what is an ideal HL simulator.
The previous version of WHIS (hereafter ${\rm WHIS_{v22}}$) was specifically based on the concept of ``cancellation of compression'' \,\citep{irino2013accurate,nagae2014hearing,irino2020gammachirp}, where the input sound level was increased as a function of the SPL to virtually reduce the compression in the IO function of an NH listener. This concept is similar to the expansion in the simulation of the loudness recruitment\,\citep{villchur1974simulation,zurek2007hearing,moore1993simulation}. \citet{baer1993effects} added a simulation of the bandwidth widening for HI listeners\,\citep{nejime1997simulation}. 

Although these approaches were practical for simulating specific functions,
we started with a more general assumption that an ideal HL simulator can provide an NH listener the same EPs of a specific HI listener by controlling the input sounds.
Obviously, a perfect simulation cannot be achieved because of the approximation and limitation of signal processing in GCFB
and the lack of knowledge about the dysfunction in the HI listener.
However, we assumed that the approximation is possible if certain small errors are allowed.
Then, the main issue is the degree of similarity between the EPs of the HI listener and simulated EPs, as described in the next section. 
The schematic IO function shown in Fig.\,\ref{fig:IOfunc_ResultGCFB} is found to be useful when considering this approach.

\begin{figure*}[t]
\centerline{\includegraphics[width=1\linewidth, bb=0 0 1047 617] {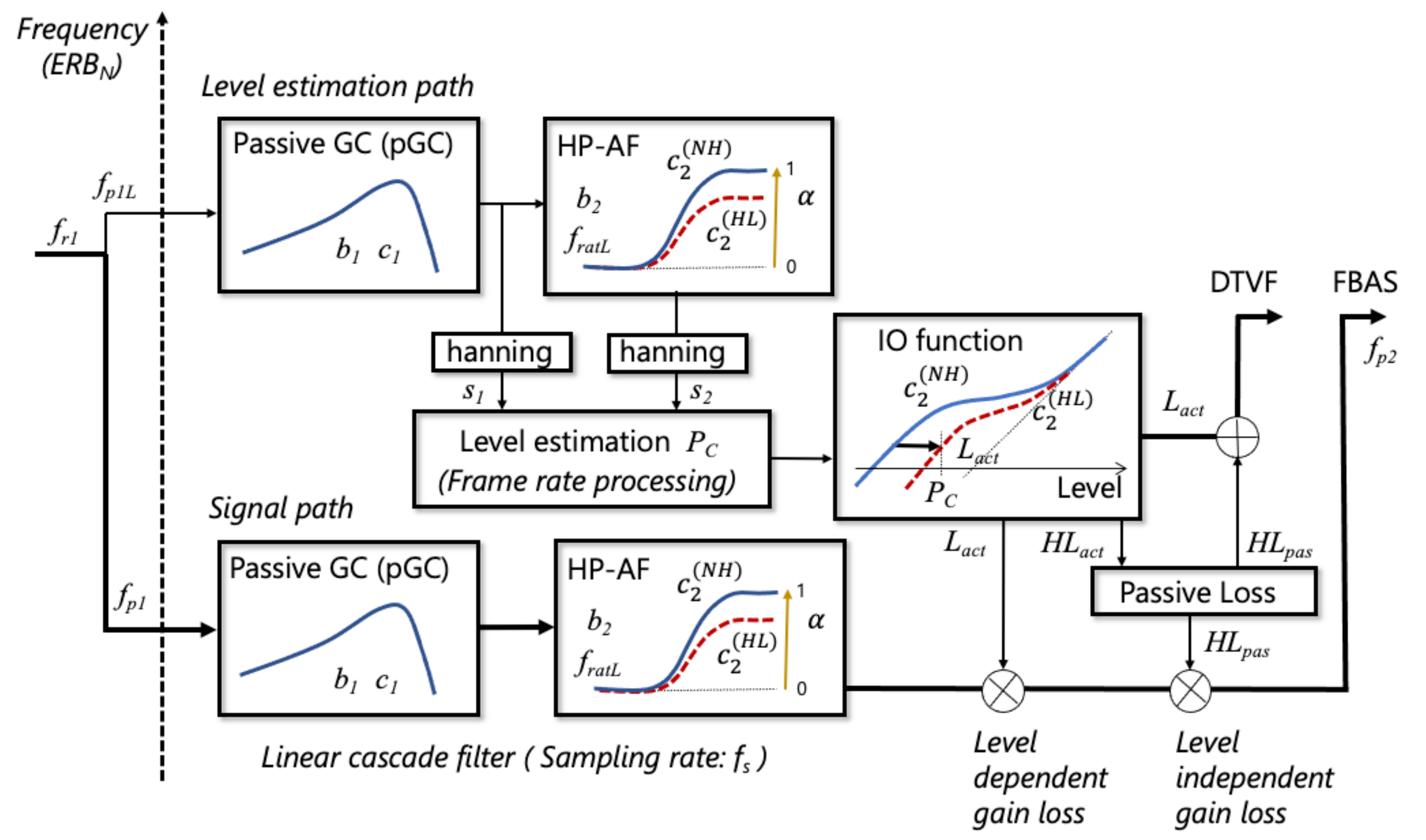}}
   \caption{Block diagram of one channel of analysis section of ${\rm WHIS_{v30}}$. Notice that the blocks in the left half is the same as those in ${\rm GCFB_{v23}}$ (Fig.\,\ref{fig:Diagram1ch_GCFBv23}) and the main difference is the use of the IO function instead of the active gain function. 
   }
\label{fig:Diagram1ch_WHISv30}
\vspace{-10pt}
\end{figure*}

\subsection{Analysis part}
\label{sec:WHISv30_Analysis}
Figure \ref{fig:Diagram1ch_WHISv30} shows a block diagram for one channel of the analysis section of ${\rm WHIS_{v30}}$. 
The blocks of the linear cascade filters (pGC + HP-AF) and the level-estimation circuit in the left half are exactly the same as those in ${\rm GCFB_{v23}}$ as shown in Fig.\, \ref{fig:Diagram1ch_GCFBv23}. 
In practice, the same software is used in this part.
The main difference lies in the use of the IO function, as shown in Fig.\,\ref{fig:IOfunc_ResultGCFB}, instead of the active gain function. The IO function is used to estimate the input level from the EP inversely. The EPs of the NH and HI listeners can be calculated by ${\rm GCFB_{v23}}$ under the NH and HI settings. We assumed that the difference between them corresponds to what WHIS tries to compensate for by converting the input sound.


The IO function, $F_{IO}$, can be defined as follows:
\begin{eqnarray}
 P_{out} & = F_{IO}(P_{in}) & = G_{act}(P_{in})+P_{in}, \label{eq:PoutFIOPin}
\end{eqnarray}
where  $P_{in}$ and $P_{out}$ are the input and output levels on a dB scale. $G_{act}$ is the active gain, on a dB scale, which is shown in Fig.\,\ref{fig:IOfunc_schematic} and defined in Eq. \ref{eq:Gact}.  The conversion from the EP (i.e., the output level) to the input sound level is performed by using the inverse IO function, defined as 
\begin{eqnarray}
 P_{in} =  F_{IO}^{-1}(P_{out}). 
 \label{eq:PinFIO-1}
\end{eqnarray}

The algorithms for calculating the signal levels for the simulated HL are formulated as the followings.
The passive HL is initially ignored for simplicity, as shown in  Fig.\,\ref{fig:IOfunc_schematic}.
The input level necessary to achieve a certain target output level $P_{out}^{(target)}$ (e.g., 10\,dB in Fig.\,\ref{fig:IOfunc_schematic}) for an HI listener can be formulated as:
\begin{eqnarray}
  P_{out}^{(target)} & = & F_{IO}^{(HL)}(P_{in}^{(HL)}),
    \label{eq:PoutTargetHL} 
\end{eqnarray}
where $F_{IO}^{(HL)}$ is the IO function shown by the red dashed line in Fig.\,\ref{fig:IOfunc_schematic}, and $P_{in}^{(HL)}$ is the input level for the HI listener.
The same output level $P_{out}^{(target)}$ is achieved by an NH listener as
\begin{eqnarray}
  P_{out}^{(target)} & = & F_{IO}^{(NH)}(P_{in}^{(NH)}),
    \label{eq:PoutTargetNH} 
\end{eqnarray}
where $F_{IO}^{(NH)}$ is the IO function shown by the blue solid line in Fig.\,\ref{fig:IOfunc_schematic}, and $P_{in}^{(NH)}$ is the input level for the NH listener.
As described previously,
the objective of WHIS is to equalize the EPs of the HI and NH listeners. 
The difference between $P_{in}^{(NH)}$ and $P_{in}^{(HL)}$
can be interpreted as active loss, $L_{act}$, in Fig.\,\ref{fig:IOfunc_schematic} in the cochlear input as
\begin{eqnarray}
  L_{act} & = & P_{in}^{(HL)} - P_{in}^{(NH)}. 
 \end{eqnarray}
This equation can be rewritten using in Eqs.\,\ref{eq:PinFIO-1} --  \ref{eq:PoutTargetNH} as
\begin{eqnarray}
   L_{act} & = & F_{IO}^{(HL)^{-1}}(P_{out}^{(target)}) -  F_{IO}^{(NH)^{-1}}(P_{out}^{(target)}).
  \label{eq:LactFIO}
\end{eqnarray}
Therefore, the active loss, $L_{act}$, is calculated at the same output level, $P_{out}^{(target)}$,
and represented as a horizontal shift, as shown in Fig.\,\ref{fig:IOfunc_schematic}.

The WHIS circuit shown in Fig. \ref{fig:Diagram1ch_WHISv30}, $P_{out}^{(target)}$ is initially calculated by Eq. \ref{eq:PoutTargetHL} using the frame-based level, $P_{C}(\tau)$, estimated by the level estimation circuit as
\begin{eqnarray}
   P_{out}^{(target)} & = & F_{IO}^{(HL)}(P_{c}(\tau)).
\end{eqnarray}
When substituting this equation into Eq.\,\ref{eq:LactFIO}, the frame-based active loss, $L_{act}(P_c(\tau))$,  can be derived as
\begin{eqnarray}
   L_{act}(P_c(\tau)) & = & F_{IO}^{(HL)^{-1}}\{F_{IO}^{(HL)}(P_{c}(\tau))\} -  F_{IO}^{(NH)^{-1}}\{F_{IO}^{(HL)}(P_{c}(\tau))\} \nonumber \\
   & = & P_{c}(\tau) -  F_{IO}^{(NH)^{-1}}F_{IO}^{(HL)}(P_{c}(\tau)).
\end{eqnarray}
Consequently, the active loss, $L_{act}(P_c(\tau))$, can be simply determined using the composite function, which comprises the IO function of HI,  $F_{IO}^{(HL)}$, and the inverse IO function of NH, $F_{IO}^{(NH)^{-1}}$. 
$L_{act}$ becames the same as $HL_{act}$ in Eq. \ref{eq:HLtotal_ACT+PAS} 
when the output level, $P_{out}$, is zero (i.e., at the AT level), as shown in Fig.\,\ref{fig:IOfunc_schematic}.
Therefore, $HL_{act}$ can be calculated directly from Eq.\,\ref{eq:LactFIO} when $P_c(\tau)$ is set to the AT of the HL listener.
Once $HL_{act}$ is determined, the passive loss, $HL_{pas}$, can be easily determined by using Eq. \ref{eq:HLtotal_ACT+PAS} as
\begin{eqnarray}
  HL_{act} & = & L_{act}\,|\,_{P_{out} =0}\\
  HL_{pas} & = & HL_{total} - HL_{act}
\end{eqnarray}
The total loss for the HL simulation is derived from these equations as follows:
\begin{eqnarray}
   L_{total}(P_c(\tau))  & = &  L_{act}(P_c(\tau)) + HL_{pas}.
   \label{eq:WHIS_LossTotal_nonch}
\end{eqnarray}
Note that the signal processing is performed in each filterbank channel $n_{ch}$ with the estimated signal level $P_c(n_{ch}, \tau)$.
Equation \,\ref{eq:WHIS_LossTotal_nonch} can be rewritten as follows:
\begin{eqnarray}
   L_{total}(n_{ch},P_c(n_{ch},\tau))  & = &  L_{act}(n_{ch},P_c(n_{ch},\tau)) + HL_{pas}(n_{ch}).
   \label{eq:WHIS_LossTotal}
\end{eqnarray}
Using this equation, the input sound level necessary to simulate the HL can be determined.
This algorithm is simpler and more intuitive than that in the previous WHIS \,\citep{irino2013accurate,nagae2014hearing,irino2020gammachirp}.
The IO function is operated horizontally as shown in Fig.\,\ref{fig:IOfunc_schematic}. Although this equation is similar to Eq. \ref{eq:Gtotal_nch_Pc=Gact}, the IO function is operated vertically in GCFB. 
This difference poses the fundamental limit of HL simulators, as described later.

The analysis part in Fig.\,\ref{fig:Diagram1ch_WHISv30} produces two types of outputs, which correspond to the synthesis methods described in the next section.
One of them is the frame-based loss, $L_{total}(n_{ch},P_c(\tau))$, described in Eq. \ref{eq:WHIS_LossTotal}, itself. The other one is the cascade filter output, where the amplitude is dynamically reduced by the resampled version of the total loss $L_{total}(n_{ch},P_c(n_{ch},\tau))$.


\subsection{Synthesis part}
\label{sec:WHISv30_Synthesis}
The simulated HL sounds are synthesized from the analysis part by using two methods.

\begin{figure*}[t]
  \begin{center} 
    \includegraphics[width=1\linewidth]{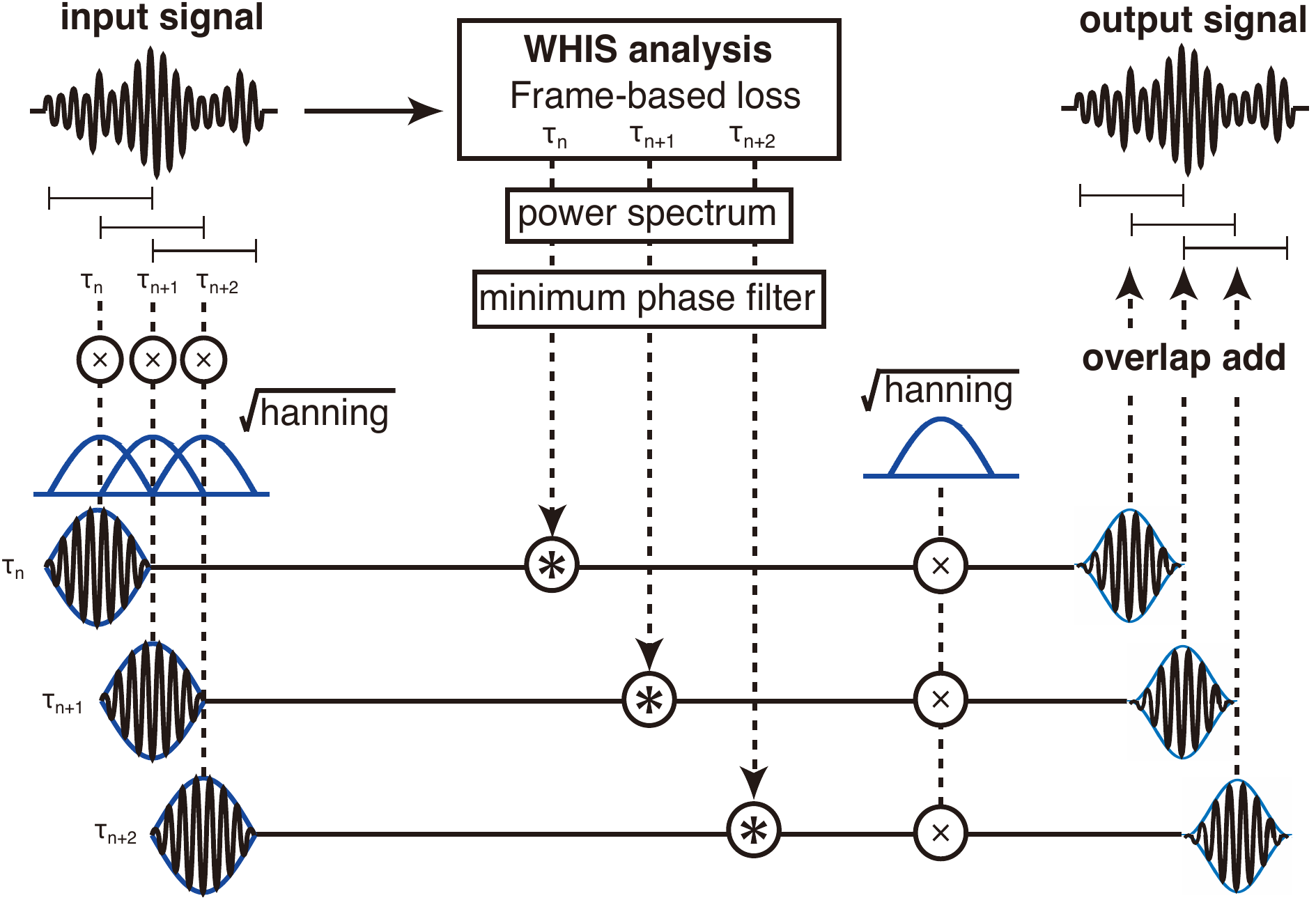}
    \caption{Signal synthesis using a DTVF.} 
    \vspace{-3mm}
    \label{fig:TimeVaryingFilter} 
  \end{center}
\end{figure*}

\subsubsection{Direct time-varying filter}
\label{sec:DTVF}
One of the synthesis methods is to apply a nonlinear, time-varying filter to input signals, as used in the previous WHIS\,\citep{irino2020gammachirp}.
This is referred to as direct time-varying filtering (DTVF) hereafter. The filter coefficients are calculated from the frame-based loss $L_{total}(n_{ch},P_c(\tau))$ in Eq.\,\ref{eq:WHIS_LossTotal}.
Figure \ref{fig:TimeVaryingFilter} presents a schematic of signal processing. 
The input signal is divided into frames with a square-root hanning window, $ w(t) = \sqrt{0.5 + 0.5 \cos(2 \pi t/T)} ~\{t |-T/2 \leq t \leq T/2\}$, where the frame length, $T$, is 20 ms and the frame shift is 10 ms.  The framed signal is convoluted with a minimum phase filter, which is described in the next paragraph. The filtered signal is windowed again with the same square-root hanning window, $w(t)$. Then, the frames are then overlap-and-added to produce the output signal. When the minimum phase filter is an impulse, the input and output signals are identical, since this procedure is equivalent to processing with a hanning window with half overlapping. 

The minimum phase filter is derived from the output of the path labeled ``DTVF'' in Fig.\,\ref{fig:Diagram1ch_WHISv30}. 
The frame-based loss $L_{total}(n_{ch}, P_c(\tau))$ in Eq.\,\ref{eq:WHIS_LossTotal} is interpreted as the spectral distribution of the loss function along the filter channel, $n_{ch}$ (i.e., on the $\rm{ERB_Nnumber}$ axis). This distribution is converted into the power spectrum on the linear frequency axis by using a warping function from $\rm{ERB_N}number$ to Hz. Then, the minimum phase filter is derived from this power spectrum using the cepstral method.

WHIS with this DTVF synthesis method is referred to as ${\rm WHIS_{v30}^{DTVF}}$.
The results of the preliminary listening tests indicate that ${\rm WHIS_{v30}^{DTVF}}$ does not produce noticeable distortion in the output sounds, which is 
attributable to a single time-varying filter between the input and output for each frame. The filter has a minimum phase response that does not produce pre-echo, which might be perceived as distortion.

However, there is also a disadvantage.
It is difficult to introduce the process of temporal smearing because the analysis output  
is the frame-based loss $L_{total}(n_{ch},P_c(\tau))$ rather than the EP, which contains the temporal envelope.
Although an additional analysis-synthesis filterbank would be applicable to the output sound as post-processing, it becomes inconsistent with the aimed unified framework and results in increasing the distortion.


\subsubsection{Filterbank synthesis}
\label{sec:WHISv30fabs}
The filterbank synthesis is an alternative method to avoid the disadvantage of DTVF. The output sound is synthesized by an overlap-and-add method, which is commonly used and is similar to that in \citet{irino1999analysis, irino2006dynamic}. In the current implementation for fast processing, the phase delay of the output waveform from the individual filterbank channel is compensated for a constant reciprocal to the center frequency of the corresponding gammachirp filter. Then, the compensated waveforms are added together to synthesize the output.
The WHIS obtained with this filterbank analysis-synthesis (FBAS) method is referred to as ${\rm WHIS_{v30}^{FBAS}}$.
The process in the single channel is shown in the path labeled ``FBAS'' in Fig.\,\ref{fig:Diagram1ch_WHISv30}. The amplitude of the output waveform of the linear cascade filter is reduced by $L_{act}(n_{ch},Pc(\tau))$ and $HL_{pas}(n_{ch})$, as defined in Eq.\,\ref{eq:WHIS_LossTotal}. The process is performed with adequate resampling from the frame rate to the signal sampling rate.

${\rm WHIS_{v30}^{FBAS}}$ can accommodate the temporal smearing method within a single framework. It is similar to the method used by \,\citet{drullman1994effect}. The envelope is extracted from the filterbank output by Hilbert transformation or rectification and is filtered using a lowpass filter designed to reduce the temporal resolution. The original carrier component and the reduced envelope are used to synthesize the output sound. 

In the preliminary listening tests, the output sounds in ${\rm WHIS_{v30}^{FBAS}}$ are slightly distorted, even without any temporal smearing. The distortion level is slightly higher than that in ${\rm WHIS_{v30}^{DTVF}}$. 
The phase compensation across the filter channels is not perfectly performed, probably because the temporal response of the cascade filter (pGC + HP-AF) in Fig.\ref{fig:Diagram1ch_WHISv30} is determined not only by the center frequency but also by the compression health $\alpha$.  Thus, to achieve better quality, more sophisticated processing is required.

\section{Evaluation of WHIS}
\label{sec:WHIS_Evaluation}
We evaluate the simulated HL sounds of speech to clarify the potential and limit of the following four HL simulators: ${\rm WHIS_{v30}^{DTVF}}$, ${\rm WHIS_{v30}^{FBAS}}$,  ${\rm WHIS_{v22}}$ (i.e. the previous version of WHIS), and CamHLS. 
The HL sounds simulated for an average 80-years-old hearing level\,\citep{tsuiki2002nihon} are analyzed using ${\rm GCFB_{v23}}$ under the NH setting. For convenience, this process is referred to as ``WH-GC(NH)'' hereafter. 
We also analyze the original sounds by using ${\rm GCFB_{v23}}$ with the same 80-years-old setting. This process is referred to as ``GC(HL).''
The output representations of WH-GC(NH) and GC(HL) are compared to evaluate the goodness of the simulation. 
The difference between WH-GC(NH) and GC(HL) is assumed to be greater in the 80-years-old hearing level than in a milder hearing level.

\begin{figure*}[t]
  \centerline{\includegraphics[width=1.1\linewidth]{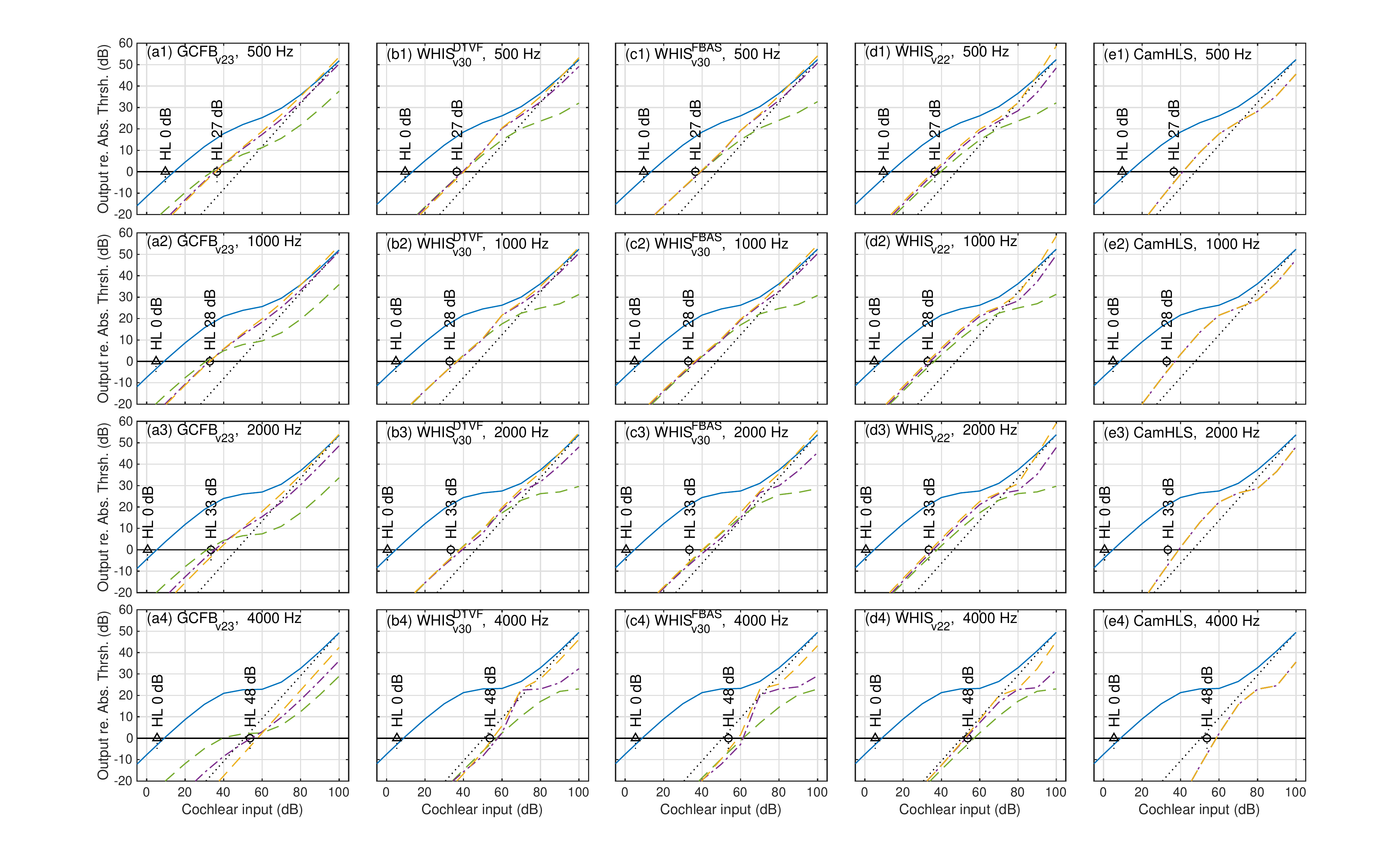}}
  \caption{Simulation results on the IO functions at  frequencies of  500, 1000, 2000, and 4000\,Hz. The axes of the figures, the plotted lines, and the HL values are the same as those in Fig.\,\ref{fig:IOfunc_ResultGCFB}. The panels of (a1)-(a4) show the IO functions of GC(HL), which are the same as those shown in Fig.\,\ref{fig:IOfunc_ResultGCFB} for corresponding frequencies. The other panels show the IO functions of WH-GC(NH) with  ${\rm WHIS_{v30}^{DTVF}}$ ((b1)-(b4)) , with  ${\rm WHIS_{v30}^{FABS}}$ ((c1)-(c4)) , with ${\rm WHIS_{v22}}$ ((d1)-(d4)), and with CamHLS ((e1)-(e4)). Blue solid line: Average NH. 
Green dashed line: HL with the compression health $\alpha$ of 1. Purple dashed-and-dotted line: 
HL with $\alpha$ of 0.5. Orange dashed line: HL with $\alpha$ of 0. Black dotted line: The linear relationship (1:1).
}
\label{fig:IOfunc_ResultGCFBWHISCamHLS}

\end{figure*}

\subsection{IO function}
\label{sec:WHIS_IOfunction}
The IO functions are calculated by using short sinusoids, which were used for producing Fig.\,\ref{fig:IOfunc_ResultGCFB}.
Figure\,\ref{fig:IOfunc_ResultGCFBWHISCamHLS} shows the simulation results.
The panels of (a1)-(a4) show the IO functions of GC(HL) at the frequencies of  500, 1000, 2000, and 4000\,Hz. These panels are identical to those shown in Fig.\,\ref{fig:IOfunc_ResultGCFB} for the corresponding frequencies.
The IO functions in panels (b1)-(b4) are for WH-GC(NH) with ${\rm WHIS_{v30}^{DTVF}}$; those in (c1)-(c4) are for WH-GC(NH) with  ${\rm WHIS_{v30}^{FABS}}$; those in (d1)-(d4) are for WH-GC(NH) with ${\rm WHIS_{v22}}$; and those in (e1)-(e4) are for  WH-GC(NH) with CamHLS.

\subsubsection{Comparison with GCFB}
\label{sec:WHIS_IOfunction_CompareGCFB}

First, let us compare panels (a1)-(a4) for GC(HL) and (b1)-(b4) for WH-GC(NH) with ${\rm WHIS_{v30}^{DTVF}}$.
The IO functions of the NH conditions (blue solid lines) are clearly the same at the same frequency because no HL processing is performed in WH-GC(NH). The IO functions for $\alpha$ of 0.0 and 0.5 (orange and purple lines, respectively) are similar except for 4000\,Hz. The major differences are observed at $\alpha = 1.0$ (green dashed lines). The locations of the compressive regions are higher in WH-GC(NH) than in GC(HL), which can be explained by the difference in the gain control.
The output levels in panels (a1)-(a4) are calculated from $G_{total}(n_{ch},P_c(\tau))$ in Eq.\,\ref{eq:Gtotal_nch_Pc=Gact}, which operates the IO functions vertically, as shown in Fig.\,\ref{fig:IOfunc_schematic}. At $\alpha = 1.0$, the NH IO functions (blue line) are shifted downward for $L_{pas}(n_{ch})$ to the HL IO functions (green dashed line).
In contrast, the operations in any HL simulators are completely different because they solely control the input signal levels and cannot touch the output levels directly. The input levels in ${\rm WHIS_{v30}^{DTVF}}$ are controlled by $L_{total}(n_{ch},P_c(\tau))$ in Eq.\,\ref{eq:WHIS_LossTotal}, which operates the IO functions horizontally, as shown in Fig.\,\ref{fig:IOfunc_schematic}. At $\alpha = 1.0$, the NH IO functions (blue solid line) are shifted rightward for $HL_{pas}(n_{ch})$ to the HL IO functions (green dashed line). As a result, the IO functions are different in panels (a1)-(a4) and (b1)-(b4).
The degree of inconsistency is smaller at $\alpha \le 0.5$. This is because, at this value, the IO functions are less compressive, and the vertical shift induced by the passive loss $L_{pas}(n_{ch})$ in Eq.\,\ref{eq:Gtotal_nch_Pc=Gact} and the horizontal shift induced by the passive loss $HL_{pas}(n_{ch})$ in Eq.\, \ref{eq:WHIS_LossTotal} are relatively smaller as compared to the case of $\alpha=1$.
The results reveal the fundamental limit of any existing HL simulator.

\subsubsection{Comparison between the HL simulators}
\label{sec:WHIS_IOfunction_CompareHLS}

The IO functions in panels (c1)-(c4) for WH-GC(NH) with ${\rm WHIS_{v30}^{FBAS}}$ in Fig. \ref{fig:IOfunc_ResultGCFBWHISCamHLS} are the same as those in panels (b1)-(b4) for WH-GC(NH) with ${\rm WHIS_{v30}^{DTVF}}$.
This means that the synthesis parts in ${\rm WHIS_{v30}^{DTVF}}$ and ${\rm WHIS_{v30}^{FBAS}}$ do not affect the results when the sinusoids are processed.

 The differences between panels (d1)-(d4) for WH-GC(NH) with ${\rm WHIS_{v22}}$ and panels (b1)-(b4) for WH-GC(NH) with ${\rm WHIS_{v30}^{DTVF}}$ are mainly observed at $\alpha=0.5$, where the input levels exceed 80\,dB. However, the IO functions at $\alpha=1.0$ and $\alpha=0.0$ are very similar. The difference at $\alpha=0.5$ is attributable to that in the definitions of $\alpha$ in ${\rm WHIS_{v30}}$  (Eq.\,\ref{eq:c2HL=alpha}) and in ${\rm WHIS_{v22}}$ \,\citep{irino2020gammachirp}, although the range is the same $\{\alpha | 0.0 \le \alpha \le 1.0\}$. Therefore, the output of ${\rm WHIS_{v22}}$ might be similar to that of ${\rm WHIS_{v30}}$ when the $\alpha$ value is adequately converted. The results imply that the perceptual experiments performed using ${\rm WHIS_{v22}}$ can be
interpreted consistently with those performed using ${\rm WHIS_{v30}}$.
 
There is a single IO function in WH-GC(NH) with CamHLS for each frequency, as shown in panels (e1)-(e4). Under the default setting, the IO function as automatically determined from the given audiogram. The IO functions are similar to those in ${\rm WHIS_{v30}}$ at $\alpha=0.5$. This means that both ${\rm WHIS_{v30}}$ and CamHLS can simulate the loudness recruitment.


\subsection{Distance between auditory spectrograms}
\label{sec:SpecDistance}
We evaluate the HL simulators using speech sounds because such nonlinear systems cannot be evaluated sufficiently by simple sinusoids. The auditory spectrograms calculated from ${\rm GCFB_{v23}}$ are used for the evaluation. 
\subsubsection{Method}
\label{sec:SpecDistance_Method}

The speech sounds are analyzed using the GC(HL) process, as described above,
to derive the reference auditory spectrogram $S(n_{ch},\tau)_{GC(HL)}$, where $n_{ch}$ is the filterbank channel and $\tau$ is the frame time. The frame window length is 1.0\,ms,  and the frame shift is 0.5\,ms. The input speech sounds are normalized at the SPLs of 50 and 80\,dB in $L_{eq}$ (i.e., rms level). The $\alpha$ values are set to 1.0, 0.5, and 0.0. 
For the analysis, 20 speech sounds pronounced by two male and two female speakers are drawn from the Japanese word database FW07 \,\citep{FW07}.

The auditory spectrograms of the simulated HL sounds  $S(n_{ch},\tau)_{GC(NH)}^{WH}$ are calculated by the WH-GC(NH) process.
The spectral distance between $S(n_{ch},\tau)_{GC(NH)}^{WH}$ and $S(n_{ch},\tau)_{GC(HL)}$ is used for a goodness measure of the HL simulation. The simulation result is good enough if the distance is very small.
We defined the normalized spectral distance $d_{sp}$ (dB) as
\begin{eqnarray}
 d_{sp} &=& 10\log_{10}\biggl[ \frac{ \sum_{n_{ch}}\sum_{\tau}\{S(n_{ch},\tau+\Delta\tau)_{GC(NH)}^{WH}- S(n_{ch},\tau)_{GC(HL)}\}^2} {\sum_{n_{ch}}\sum_{\tau}{S(n_{ch},\tau)_{GC(HL)}}^2} \biggr].
 \label{eq:ds=10log10}
\end{eqnarray}
where $\Delta\tau$ is the frame shift which yields the minimum distance. This is necessary because the start time of the HL-simulated sound might be different from that of the original sound. $\Delta \tau$ is searched within a limited range. 

Moreover, noisy sounds are analyzed using the GC(HL) process to estimate the degree of distortion. Pink noise is added to the original speech sound with SNRs of +3, 0, and -3\,dB. The derived auditory spectrogram is substituted for $S(n_{ch},\tau)_{GC(NH)}^{WH}$ in Eq. \ref{eq:ds=10log10}.
Therefore, in this case, the $d_{sp}$ values are calculated between the noisy and clean speech sounds.
The $d_{sp}$ values can help interpret the results, although the nonlinear distortion and the additive noise are completely different in perceptual impression.


\begin{figure*}[t]
\vspace{-10pt}
\centerline{\includegraphics[width=1\linewidth]{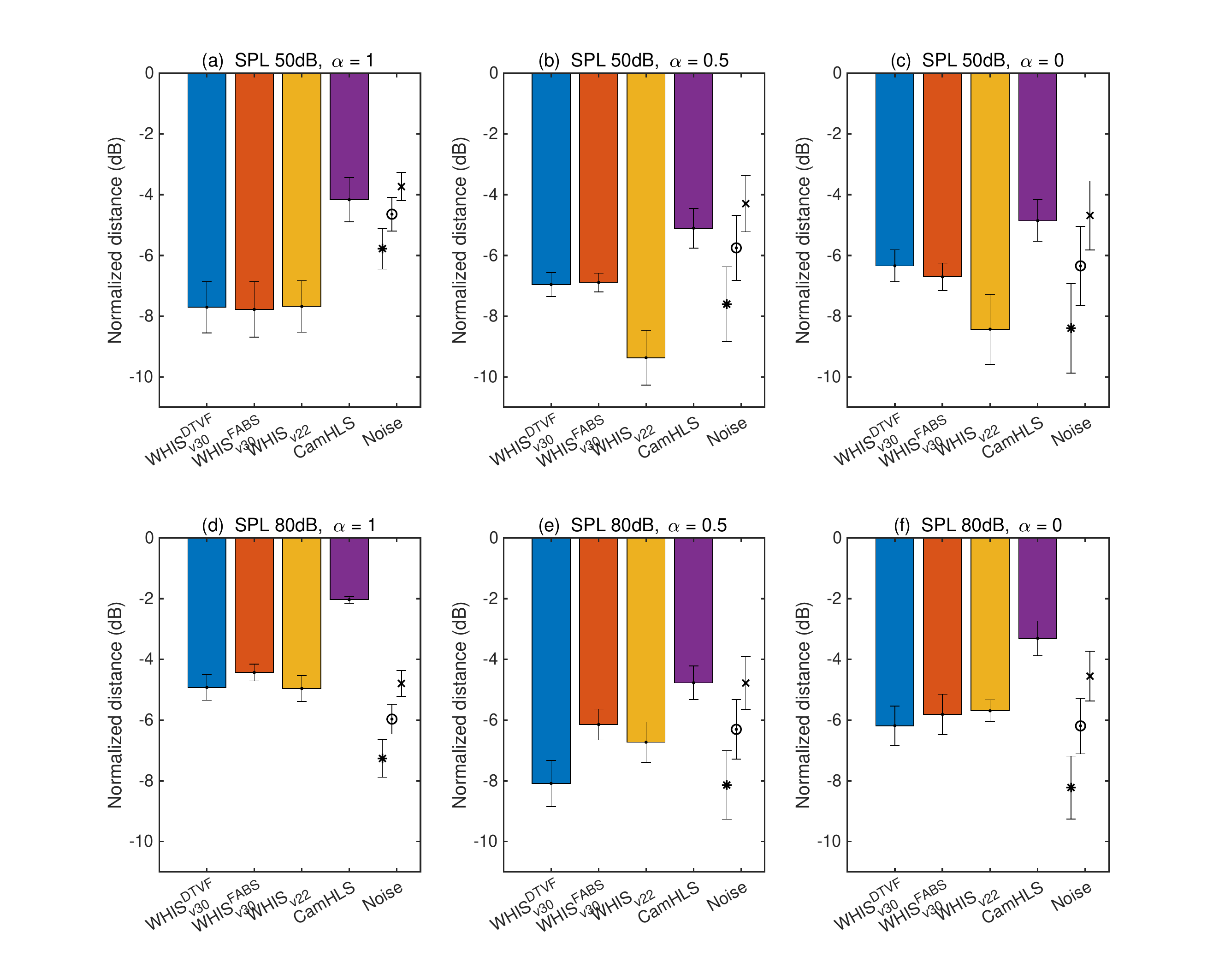}}
  \vspace{-5pt}
  \caption{Spectral distance $d_{sp}$ in Eq.  \ref{eq:ds=10log10}. Bars and error bars represent the mean and standard deviation (SD) of $d_{sp}$. The conditions for the SPL and $\alpha$ are labeled at the top of each panel. The HL simulator conditions are ${\rm WHIS_{v30}^{DTVF}}$ (blue), ${\rm  WHIS_{v30}^{FBAS}}$ (red), ${\rm WHIS_{v22}}$ (orange), and CamHLS (purple). The $d_{sp}$ values for the noisy sounds when the SNRs are 3\,dB(*), 0\,dB(o), and -3\,dB ($\times$) are also plotted in the rightmost part. }  
\label{fig:SpecDistance}
\vspace{-10pt}
\end{figure*}


\subsubsection{Results}
\label{sec:SpecDistance_Result}
Figure \ref{fig:SpecDistance} shows the mean and standard deviation (SD) of the $d_{sp}$ values. We compare ${\rm WHIS_{v30}^{DTVF}}$, ${\rm WHIS_{v30}^{FBAS}}$, and  ${\rm WHIS_{v22}}$ in the WHIS family and CamHLS.  The $d_{sp}$ values of noisy sounds at the SNRs of 3, 0, and -3\,dB are also plotted on the right to obtain information about the degree of the distortion. 

\paragraph{ANOVA}
We conduct a three-factor ANOVA on the HL simulation condition, the SPL, and the $\alpha$ value. The results show significant differences in the main effect and in interaction ($p<10^{-10}$ under any condition). We then analyze the differences in the individual factors.

\paragraph{Difference between the SPLs and the $\alpha$ conditions}
The spectral distances in the WHIS family at $\alpha = 1$ were approximately 3\,dB greater under  80\,dB condition (panel (d)) than under  50\,dB condition  (panel (a)). 
This is probably because the IO functions at $\alpha = 1$ are largely different between the GC(HL) and WH-GC(NH) conditions at a high SPL,
as shown in Fig.\,\ref{fig:IOfunc_ResultGCFBWHISCamHLS}.
There are various differences between the two SPL conditions at $\alpha = 0$ and $\alpha = 0.5$.

\paragraph{Difference between the WHIS family and CamHLS}
The spectral distances are always greater in CamHLS than in the WHIS family, as observed in the all panels. This is probably because spectral smearing is introduced into CamHLS to simulate the bandwidth widening in the HI listeners \,\citep{baer1993effects,baer1994effects}. The process of the spectral smearing produces distortion, which is described as
{\it ``The `spectral smearing' software has the perceptual effect of adding a certain `grittiness' to the audio quality''} in the software (control\_impaired\_ear.m).
This distortion can be avoided and the spectral distance can be reduced by removing the spectral smearing process.

\paragraph{Difference within the WHIS family}
The distances are smaller in 
${\rm WHIS_{v22}}$ than in ${\rm WHIS_{v30}^{DTVF}}$ and ${\rm WHIS_{v30}^{FBAS}}$ under the 50\,dB condition and at $\alpha = 0.5$ and $\alpha = 0$ (panels (b) and (c)).
In contrast, the distance is the smallest in ${\rm WHIS_{v30}^{DTVF}}$ under the 80\,dB condition and at  $\alpha=0.5$ (panel (e)). In the other conditions, the distances in ${\rm WHIS_{v30}^{DTVF}}$ and ${\rm WHIS_{v30}^{FBAS}}$ are approximately the same .

One of the purposes of the HL simulator is to provide insight into how loud the sounds are required for the HI listeners. ${\rm WHIS_{v30}^{DTVF}}$ might be advantageous for this purpose because the distortion is the minimum under the 80\,dB condition. However, the differences among ${\rm WHIS_{v30}^{DTVF}}$, ${\rm WHIS_{v30}^{FBAS}}$, and ${\rm WHIS_{v22}}$ are small; they seem to be compatible in general.

\paragraph{Difference between the HL-simulated sounds and the noisy sounds}
It is not easy to interpret whether the value of the spectral distance is sufficiently small. We calculate the spectral distances between the clean and noisy sounds at SNRs of 3,  0, and -3\,dB under the GC(HL) condition. The results are shown in the rightmost part in each panel.
The spectral distances of the WHIS family are generally smaller than or equal to those of the noisy sounds at  0\,dB SNR. In the case of panel (d), the spectral distance is almost the same as that of the noisy sounds at -3\,dB SNR. 
Therefore, the distances are not very small. However, the distortion and noise components in the sounds simulated by WHIS are not perceptually salient, unlike the additive noise sounds. WHIS does not produce any ``grittiness'' component.  WHIS accomplishes the HL simulation up to this quality.  


\section{Discussion}
\label{sec:Discussion}

\subsection{Fundamental limit of the HL simulators}
\label{sec:FundLimit_HLsimulator}

The results of the IO function shown in Fig.\,\ref{fig:IOfunc_ResultGCFBWHISCamHLS} and the spectral distance shown in Fig.\,\ref{fig:SpecDistance} demonstrate that the HL simulation cannot be accurately performed when $\alpha$ is close to 1, or in other words, when the active process including the OHC function is healthy. 
This implies that it is difficult to simulate hidden HL\,\citep{liberman2015hidden}, including synaptopathy\,\citep{sergeyenko2013age}, when the OHC function is diagnosed as healthy.  This is a fundamental limit of any existing HL simulator as well as WHIS.  To overcome this problem, it is essential to simulate the compression just above the output level of 0\,dB, as shown in Fig.\,\ref{fig:IOfunc_ResultGCFB} and in Fig.\,\ref{fig:IOfunc_ResultGCFBWHISCamHLS} (a1)-(a4).
However, this is not easy to achieve and remains problematic.

In contrast, the HLs of many elderly HI listeners might be caused by the dysfunctions in both the active and passive processes. At $\alpha$ less than 1, the HL simulation improves, as shown in Fig.\,\ref{fig:IOfunc_ResultGCFBWHISCamHLS}.
In any case, it is essential to estimate the ratio between the active and passive HLs, as discussed in the next section.

The results also demonstrate that passive devices, such as earplugs and graphic equalizers, could not simulate the cochlear HL.
The passive attenuation simply moves the NH IO function rightward, as shown by the $\alpha=1$ line in Fig.\,\ref{fig:IOfunc_ResultGCFBWHISCamHLS}. Nonlinear processing, such as that implemented in WHIS, is essential for cochlear HL simulation.

\subsection{Estimation of the active and passive HLs}
\label{sec:RatioHLactHLpas}

It is essential to estimate the $\alpha$ value in Eq.\,\ref{eq:c2HL=alpha} or the ratio of $HL_{act}$ and $HL_{pas}$ in Eq.\,\ref{eq:HLtotal_ACT+PAS} for reliable HL simulations.  The audiogram does not provide any information about this. The estimation of the IO function can resolve this problem.
The compression in the IO function is measured psychoacoustically by using the forward masking paradigms, such as the growth-of-masking curve method \,\citep{oxenham1997behavioral} and the temporal-masking curve method\,\citep{nelson2001new}.
An alternative method is to estimate an auditory filter using simultaneous NN masking experiments\,\citep{patterson1976auditory} with various stimulus levels.
The level dependence of the auditory filter provides a good estimate of the IO function \,\citep{irino2001compressive, baker2002auditory, patterson2003extending}.

However, these psychoacoustic methods require many measurement points for reliable estimation. They require heavy experiments that take much longer than the normal tests performed in clinical sites. Therefore, it is not very easy to estimate the parameter of a target HI listener, and a method as simple as audiometry is required to estimate the $\alpha$ value or the ratio of $HL_{act}$ and $HL_{pas}$.


\begin{figure*}[t]
  \centerline{\includegraphics[width=0.8\linewidth,bb=0 0 1048 847] {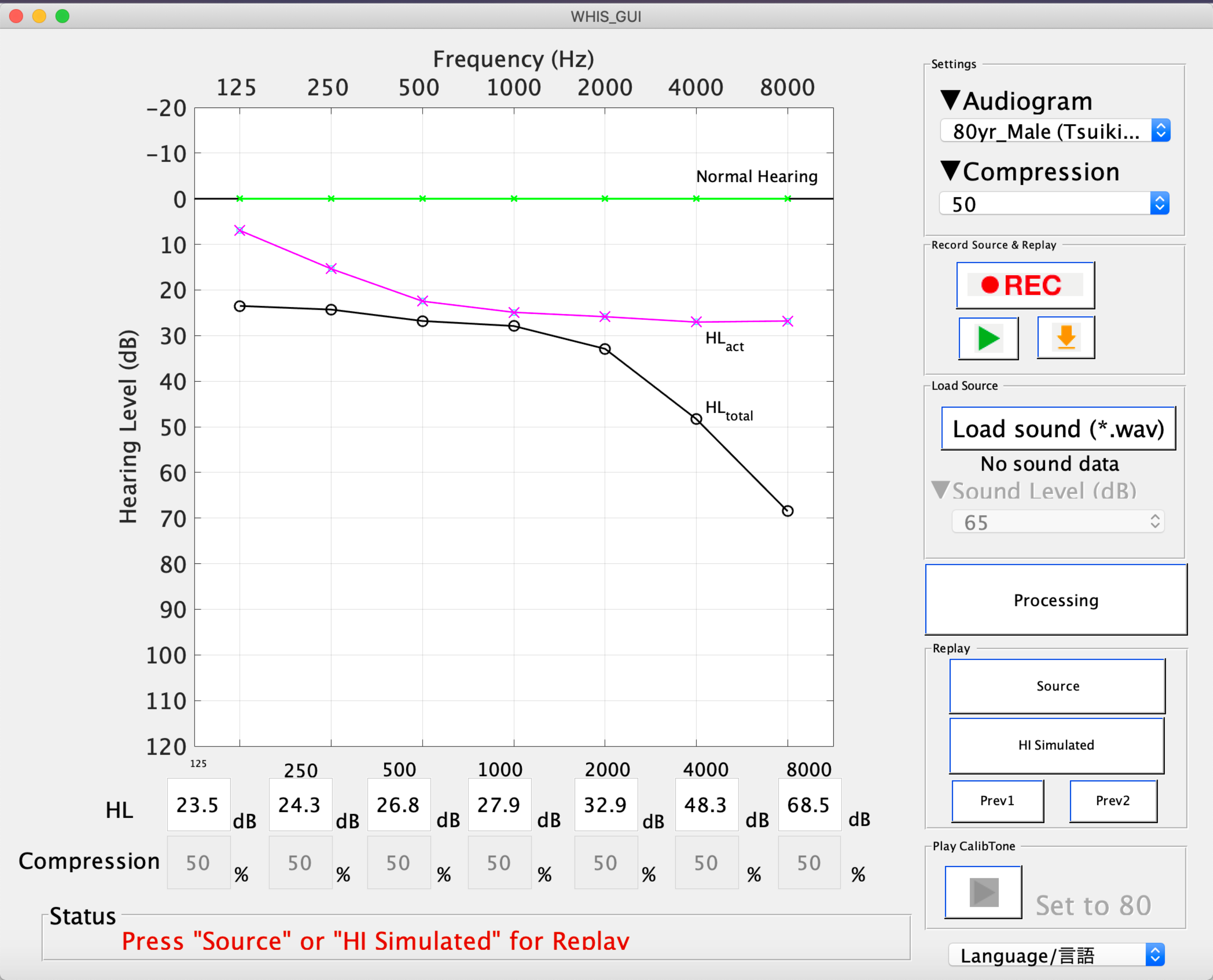}}  
 \caption{GUI of WHIS}
   \label{fig:WHIS_GUI}
 \vspace{-20pt}
\end{figure*}

\subsection{Applications of WHIS}
\label{sec:Applications}

There are many potential applications of the HL simulators.
For example, \citet{zurek2007hearing} indicated that the HL and prosthesis simulations are useful in counseling, hearing aid fitting, training, hearing conservation, testing warning signals, and setting job requirements.

An important applications includes psychoacoustic experiments using speech and environmental sounds, as indicated in the Introduction.
For this purpose, WHIS has been developed to minimize the distortion and noise, which might affect the experimental results.
Several experiments performed with WHIS are introduced in this section as examples for further studies.
For example, \citet{matsui2016effect} used an early version of WHIS to measure the effect of compression loss on syllable recognition. \citet{grimault2018real} measured the NN thresholds and estimated the auditory filter using a real-time HL simulator based on GCFB.
In addition to the perceptual experiments, \citet{irino2020vocaltrain} used ${\rm WHIS_{v22}}$ in vocal self-training experiments to determine whether the speech clarity towards HI listeners was improved.
Recently, \citet{irino2022anew} performed speech intelligibility experiments in the laboratory and crowdsourced remote environments using ${\rm WHIS_{v30}^{DTVF}}$ to clarify the effects of the listening conditions.

The stimuli in these experiments are prepared in advance using a batch program in WHIS.
A graphical user interface (GUI) shown in Fig. \ref{fig:WHIS_GUI} is used to provide NH listeners the experience of the difficulties of HI listeners interactively.
WHIS with the GUI has been used in a training program for speech-language-hearing therapists for several years \,\citep{hasegawa2019application}. 
The GUI has a main audiogram panel and several sets of control buttons. After calibration of the SPL, the user chooses an audiogram and a value of the compression health $\alpha$ in percentage. 
The audiogram is then plotted with the black line labeled with $HL_{total}$  in the main panel. In addition, $HL_{act}$ is plotted with the magenta line. At $\alpha=1$, the magenta line coincides with the green line (i.e. the hearing level of 0\,dB). The difference between the $HL_{total}$ and $HL_{act}$ lines corresponds to $HL_{pas}$, which is calculated from Eq.\,\ref{eq:HLtotal_ACT+PAS}.
Using several buttons, we can record speech sound or load prerecorded speech, and then listen to the HL-simulated sound.


\section{Conclusions }
\label{sec:Conclusions}
In this study, ${\rm WHIS_{v30}}$ was developed based on ${\rm GCFB_{v22}}$, which was updated to incorporate fast frame-based processing, the AT, the audiogram of a target HI listener, and the parameter to control the IO function. In ${\rm GCFB_{v22}}$, the compression health $\alpha$ was also introduced as an HP-AF parameter to control the degree of compression in the cochlear IO functions, which range from NH listeners to HI listeners. The total HL in the audiogram $HL_{total}$ was assumed to be the sum of the active HL $HL_{act}$
and the passive HL $HL_{pas}$ on a dB scale. 
As a result, the various IO functions were derived in accordance with $\alpha$ without much changes in the AT values. The cascade of pGC and HP-AF filters in each ${\rm GCFB_{v22}}$ channel was approximated using a linear filter to fulfill the fast processing required in WHIS. Thus, the filter shape and bandwidth were controlled by $\alpha$ but not by SPL. The estimation of $\alpha$ for the target HI listener and the introduction of the level dependence remain problems for future studies.

${\rm WHIS_{v30}}$ was developed to provide NH listeners approximately the same EPs as those of HI listeners.
The output level for the HL simulator could be controlled by a composite function of the IO function of an NH listener and the inverse IO function of a target HI listener. 
The analysis part of ${\rm WHIS_{v30}}$ was almost the same as that of ${\rm GCFB_{v22}}$, except for using the IO function instead of using the gain function. We proposed two synthesis methods: DTVF for perceptually small distortion and FBAS for additional HI simulations, including temporal smearing. 

Finally, we evaluated the WHIS family and CamHLS in terms of differences in the IO function and the spectral distance. The IO functions were well-simulated at $\alpha \le 0.5$. This was not the case at $\alpha=1$. Thus, it is  difficult to simulate the HL caused by synaptopathy when the OHC is sufficiently healthy. This is a fundamental limit of the existing HL simulator and WHIS. To overcome this problem, it is essential to precisely simulate the compression just above the AT. This also remains a problem for the future studies.
The spectral distortion was smaller in the WHIS family than in CamHLS, and ${\rm WHIS_{v30}}$ was virtually compatible with ${\rm WHIS_{v23}}$.
The software of ${\rm WHIS_{v30}}$ and ${\rm GCFB_{v22}}$ is available in our GitHub repository
\,\citep{GitHub_AMLAB}.

\section*{Acknowlegements}
This work was supported by JSPS KAKENHI Grant Numbers JP16H01734, JP18K10708, JP21H03468, and JP21K19794. The author wishes to thank Dr. Michael A. Stone for providing the CamHLS software and Dr. Yasuki Murakami for valuable comments on the earlier draft. 

\bibliographystyle{jasaauthyear2.bst}
\bibliography{Ref_WHIS_Jun22}

\begin{thebibliography}{40}
\def\enquote#1{``#1,''}
\def\plainquote#1{``#1''}
\expandafter\ifx\csname natexlab\endcsname\relax\def\natexlab#1{#1}\fi
\providecommand{\dourl}[1]{\href{http://#1}{\nolinkurl{#1}}}
\providecommand{\bibinfo}[2]{#2}
\providecommand{\noopsort}[1]{}
\providecommand{\switchargs}[2]{#2#1}
  \def\eatspace #1{#1}
\providecommand{\dodoi}[1]{doi: \href{http://dx.doi.org/#1}{\nolinkurl{#1}}}

\bibitem[{Amano \emph{et~al.}(2007)Amano, Kondo, Sakamoto, and Suzuki}]{FW07}
\bibinfo{author}{Amano, S.}, \bibinfo{author}{Kondo, T.},
  \bibinfo{author}{Sakamoto, S.},  and \bibinfo{author}{Suzuki, Y.}
  (\textbf{\bibinfo{year}{2007}}). \plainquote{\bibinfo{title}{Ntt - tohoku
  university familiarity-controlled word lists 2007 (fw07)}}
  \dourl{https://doi.org/10.32130/src.FW07}.

\bibitem[{{\rm ANSI/ASA \;S3.\;6 - 2018}(2018)}]{ansi2018specification}
\bibinfo{author}{{\rm ANSI/ASA \;S3.\;6 - 2018}}
  (\textbf{\bibinfo{year}{2018}}). \plainquote{\bibinfo{title}{Specification
  for audiometers}}
  \dourl{https://webstore.ansi.org/standards/asa/ansiasas32018}.

\bibitem[{Bacon and Viemeister(1985)}]{bacon1985temporal}
\bibinfo{author}{Bacon, S.~P.},  and \bibinfo{author}{Viemeister, N.~F.}
  (\textbf{\bibinfo{year}{1985}}). \enquote{\bibinfo{title}{Temporal modulation
  transfer functions in normal-hearing and hearing-impaired listeners}}
  \bibinfo{journal}{Audiology} \textbf{24}(2), \bibinfo{pages}{117--134},
  \dodoi{10.3109/00206098509081545}.

\bibitem[{Baer and Moore(1993)}]{baer1993effects}
\bibinfo{author}{Baer, T.},  and \bibinfo{author}{Moore, B.~C.}
  (\textbf{\bibinfo{year}{1993}}). \enquote{\bibinfo{title}{Effects of spectral
  smearing on the intelligibility of sentences in noise}} \bibinfo{journal}{The
  Journal of the Acoustical Society of America} \textbf{94}(3),
  \bibinfo{pages}{1229--1241}, \dodoi{10.1121/1.408176}.

\bibitem[{Baer and Moore(1994)}]{baer1994effects}
\bibinfo{author}{Baer, T.},  and \bibinfo{author}{Moore, B.~C.}
  (\textbf{\bibinfo{year}{1994}}). \enquote{\bibinfo{title}{Effects of spectral
  smearing on the intelligibility of sentences in the presence of interfering
  speech}} \bibinfo{journal}{The Journal of the Acoustical Society of America}
  \textbf{95}(4), \bibinfo{pages}{2277--2280}, \dodoi{10.1121/1.408640}.

\bibitem[{Baker and Rosen(2002)}]{baker2002auditory}
\bibinfo{author}{Baker, R.~J.},  and \bibinfo{author}{Rosen, S.}
  (\textbf{\bibinfo{year}{2002}}). \enquote{\bibinfo{title}{Auditory filter
  nonlinearity in mild/moderate hearing impairment}} \bibinfo{journal}{The
  Journal of the Acoustical Society of America} \textbf{111}(3),
  \bibinfo{pages}{1330--1339}, \dodoi{10.1121/1.1448516}.

\bibitem[{Clarity~Challenge(2021)}]{claritychallenge}
\bibinfo{author}{Clarity~Challenge, C.} (\textbf{\bibinfo{year}{2021}}).
  \plainquote{\bibinfo{title}{Clarity challenge}}
  \dourl{https://claritychallenge.org}, \bibinfo{note}{(Last: 21 Apr 2022)}.

\bibitem[{Drullman \emph{et~al.}(1994)Drullman, Festen, and
  Plomp}]{drullman1994effect}
\bibinfo{author}{Drullman, R.}, \bibinfo{author}{Festen, J.~M.},  and
  \bibinfo{author}{Plomp, R.} (\textbf{\bibinfo{year}{1994}}).
  \enquote{\bibinfo{title}{Effect of temporal envelope smearing on speech
  reception}} \bibinfo{journal}{The Journal of the Acoustical Society of
  America} \textbf{95}(2), \bibinfo{pages}{1053--1064},
  \dodoi{10.1121/1.408467}.

\bibitem[{Glasberg and Moore(1986)}]{glasberg1986auditory}
\bibinfo{author}{Glasberg, B.~R.},  and \bibinfo{author}{Moore, B.~C.}
  (\textbf{\bibinfo{year}{1986}}). \enquote{\bibinfo{title}{Auditory filter
  shapes in subjects with unilateral and bilateral cochlear impairments}}
  \bibinfo{journal}{The Journal of the Acoustical Society of America}
  \textbf{79}(4), \bibinfo{pages}{1020--1033}.

\bibitem[{Glasberg and Moore(2006)}]{glasberg2006prediction}
\bibinfo{author}{Glasberg, B.~R.},  and \bibinfo{author}{Moore, B.~C.}
  (\textbf{\bibinfo{year}{2006}}). \enquote{\bibinfo{title}{Prediction of
  absolute thresholds and equal-loudness contours using a modified loudness
  model}} \bibinfo{journal}{The Journal of the Acoustical Society of America}
  \textbf{120}(2), \bibinfo{pages}{585--588}, \dodoi{10.1121/1.2214151}.

\bibitem[{Grimault \emph{et~al.}(2018)Grimault, Irino, Dimachki, Corneyllie,
  Patterson, and Garcia}]{grimault2018real}
\bibinfo{author}{Grimault, N.}, \bibinfo{author}{Irino, T.},
  \bibinfo{author}{Dimachki, S.}, \bibinfo{author}{Corneyllie, A.},
  \bibinfo{author}{Patterson, R.~D.},  and \bibinfo{author}{Garcia, S.}
  (\textbf{\bibinfo{year}{2018}}). \enquote{\bibinfo{title}{A real time hearing
  loss simulator}} \bibinfo{journal}{Acta Acustica united with Acustica}
  \textbf{104}(5), \bibinfo{pages}{904--908}, \dodoi{10.3813/AAA.919252}.

\bibitem[{Hasegawa \emph{et~al.}(2019)Hasegawa, Hashi, Matsui, and
  Irino}]{hasegawa2019application}
\bibinfo{author}{Hasegawa, J.}, \bibinfo{author}{Hashi, M.},
  \bibinfo{author}{Matsui, T.},  and \bibinfo{author}{Irino, T.}
  (\textbf{\bibinfo{year}{2019}}). \enquote{\bibinfo{title}{Application of a
  hearing loss simulator to education, clinic, and research and its evaluation
  by speech-language-hearing therapists (in japanese)}} in
  \emph{\bibinfo{booktitle}{Proc. Japanese Assocication of
  Speech-Language-Hearing Therapists}}, pp. \bibinfo{pages}{1--P03--4}.

\bibitem[{Hu \emph{et~al.}(2011)Hu, Sang, Lutman, and
  Bleeck}]{hu2011simulation}
\bibinfo{author}{Hu, H.}, \bibinfo{author}{Sang, J.}, \bibinfo{author}{Lutman,
  M.~E.},  and \bibinfo{author}{Bleeck, S.} (\textbf{\bibinfo{year}{2011}}).
  \enquote{\bibinfo{title}{Simulation of hearing loss using compressive
  gammachirp auditory filters}} in \emph{\bibinfo{booktitle}{2011 IEEE
  International Conference on Acoustics, Speech and Signal Processing
  (ICASSP)}}, \bibinfo{organization}{IEEE}, pp. \bibinfo{pages}{5428--5431}.

\bibitem[{Irino \emph{et~al.}(2013)Irino, Fukawatase, Sakaguchi, Nisimura,
  Kawahara, and Patterson}]{irino2013accurate}
\bibinfo{author}{Irino, T.}, \bibinfo{author}{Fukawatase, T.},
  \bibinfo{author}{Sakaguchi, M.}, \bibinfo{author}{Nisimura, R.},
  \bibinfo{author}{Kawahara, H.},  and \bibinfo{author}{Patterson, R.~D.}
  (\textbf{\bibinfo{year}{2013}}). \enquote{\bibinfo{title}{Accurate estimation
  of compression in simultaneous masking enables the simulation of hearing
  impairment for normal-hearing listeners}} in \emph{\bibinfo{booktitle}{Basic
  Aspects of Hearing}} (\bibinfo{publisher}{Springer}), pp.
  \bibinfo{pages}{73--80}, \dodoi{10.1007/978-1-4614-1590-9_9}.

\bibitem[{Irino \emph{et~al.}(2020)Irino, Higashiyama, and
  Yoshigi}]{irino2020vocaltrain}
\bibinfo{author}{Irino, T.}, \bibinfo{author}{Higashiyama, S.},  and
  \bibinfo{author}{Yoshigi, H.} (\textbf{\bibinfo{year}{2020}}).
  \enquote{\bibinfo{title}{Speech clarity improvement by vocal self-training
  using a hearing impairment simulator and its correlation with an auditory
  modulation index.}} in \emph{\bibinfo{booktitle}{Proc. Interspeech 2020}},
  pp. \bibinfo{pages}{2507--2511}, \dodoi{10.21437/Interspeech.2020-1081}.

\bibitem[{Irino and Patterson(1997)}]{irino1997time}
\bibinfo{author}{Irino, T.},  and \bibinfo{author}{Patterson, R.~D.}
  (\textbf{\bibinfo{year}{1997}}). \enquote{\bibinfo{title}{{A time fomain,
  level-dependent auditory filter: the gammachirp}}} \bibinfo{journal}{The
  Journal of the Acoustical Society of America} \textbf{101}(1),
  \bibinfo{pages}{412--419}, \dodoi{10.1121/1.417975}.

\bibitem[{Irino and Patterson(2001)}]{irino2001compressive}
\bibinfo{author}{Irino, T.},  and \bibinfo{author}{Patterson, R.~D.}
  (\textbf{\bibinfo{year}{2001}}). \enquote{\bibinfo{title}{A compressive
  gammachirp auditory filter for both physiological and psychophysical data}}
  \bibinfo{journal}{The Journal of the Acoustical Society of America}
  \textbf{109}(5), \bibinfo{pages}{2008--2022}.

\bibitem[{Irino and Patterson(2006)}]{irino2006dynamic}
\bibinfo{author}{Irino, T.},  and \bibinfo{author}{Patterson, R.~D.}
  (\textbf{\bibinfo{year}{2006}}). \enquote{\bibinfo{title}{{A dynamic
  compressive gammachirp auditory filterbank.}}} \bibinfo{journal}{IEEE
  transactions on audio, speech, and language processing} \textbf{14}(6),
  \bibinfo{pages}{2222--2232}, \dodoi{10.1109/TASL.2006.874669}.

\bibitem[{Irino and Patterson(2020)}]{irino2020gammachirp}
\bibinfo{author}{Irino, T.},  and \bibinfo{author}{Patterson, R.~D.}
  (\textbf{\bibinfo{year}{2020}}). \enquote{\bibinfo{title}{The gammachirp
  auditory filter and its application to speech perception}}
  \bibinfo{journal}{Acoust. Sci. and Technol.} \textbf{41}(1),
  \bibinfo{pages}{99--107}, \dodoi{10.1250/ast.41.99}.

\bibitem[{Irino \emph{et~al.}(2022)Irino, Tamaru, and Yamamoto}]{irino2022anew}
\bibinfo{author}{Irino, T.}, \bibinfo{author}{Tamaru, H.},  and
  \bibinfo{author}{Yamamoto, A.} (\textbf{\bibinfo{year}{2022}}).
  \enquote{\bibinfo{title}{A new implementation of hearing impairment simulator
  whis and the effect of peripheral dysfunction on speech intelligibility (in
  japanese)}} in \emph{\bibinfo{booktitle}{Proc. Acoustical Society of Japan,
  Spring Meeting}}, pp. \bibinfo{pages}{665--668}.

\bibitem[{Irino and Unoki(1999)}]{irino1999analysis}
\bibinfo{author}{Irino, T.},  and \bibinfo{author}{Unoki, M.}
  (\textbf{\bibinfo{year}{1999}}). \enquote{\bibinfo{title}{An
  analysis/synthesis auditory filterbank based on an iir implementation of the
  gammachirp}} \bibinfo{journal}{Journal of the Acoustical Society of Japan
  (E)} \textbf{20}(6), \bibinfo{pages}{397--406}, \dodoi{/10.1250/ast.20.397}.

\bibitem[{Irino and Yamamoto(2018)}]{GitHub_AMLAB}
\bibinfo{author}{Irino, T.},  and \bibinfo{author}{Yamamoto, K.}
  (\textbf{\bibinfo{year}{2018}}). \plainquote{\bibinfo{title}{Amlab-wakayama
  github repository}} \dourl{https://github.com/AMLAB-Wakayama/},
  \bibinfo{note}{(Last: 21 Apr 2022)}.

\bibitem[{Liberman(2015)}]{liberman2015hidden}
\bibinfo{author}{Liberman, M.~C.} (\textbf{\bibinfo{year}{2015}}).
  \enquote{\bibinfo{title}{Hidden hearing loss}} \bibinfo{journal}{Scientific
  American} \textbf{313}(2), \bibinfo{pages}{48--53},
  \dourl{https://www.jstor.org/stable/26046106}.

\bibitem[{Matsui \emph{et~al.}(2016)Matsui, Irino, Nagae, Kawahara, and
  Patterson}]{matsui2016effect}
\bibinfo{author}{Matsui, T.}, \bibinfo{author}{Irino, T.},
  \bibinfo{author}{Nagae, M.}, \bibinfo{author}{Kawahara, H.},  and
  \bibinfo{author}{Patterson, R.~D.} (\textbf{\bibinfo{year}{2016}}).
  \enquote{\bibinfo{title}{The effect of peripheral compression on syllable
  perception measured with a hearing impairment simulator}} in
  \emph{\bibinfo{booktitle}{Physiology, Psychoacoustics and Cognition in Normal
  and Impaired Hearing}} (\bibinfo{publisher}{Springer, Cham}), pp.
  \bibinfo{pages}{307--314}, \dodoi{10.1007/978-3-319-25474-6}.

\bibitem[{Moore and Glasberg(1993)}]{moore1993simulation}
\bibinfo{author}{Moore, B.~C.},  and \bibinfo{author}{Glasberg, B.~R.}
  (\textbf{\bibinfo{year}{1993}}). \enquote{\bibinfo{title}{Simulation of the
  effects of loudness recruitment and threshold elevation on the
  intelligibility of speech in quiet and in a background of speech}}
  \bibinfo{journal}{The Journal of the Acoustical Society of America}
  \textbf{94}(4), \bibinfo{pages}{2050--2062}, \dodoi{10.1121/1.407478}.

\bibitem[{Moore \emph{et~al.}(1997)Moore, Glasberg, and Baer}]{moore1997model}
\bibinfo{author}{Moore, B.~C.}, \bibinfo{author}{Glasberg, B.~R.},  and
  \bibinfo{author}{Baer, T.} (\textbf{\bibinfo{year}{1997}}).
  \enquote{\bibinfo{title}{A model for the prediction of thresholds, loudness,
  and partial loudness}} \bibinfo{journal}{Journal of the Audio Engineering
  Society} \textbf{45}(4), \bibinfo{pages}{224--240},
  \dourl{http://www.aes.org/e-lib/browse.cfm?elib=10272}.

\bibitem[{Moore(2013)}]{moore2013introduction}
\bibinfo{author}{Moore, B. C.~J.} (\textbf{\bibinfo{year}{2013}}).
  \emph{\bibinfo{title}{An introduction to the psychology of hearing}},
  \bibinfo{edition}{6th} ed. (\bibinfo{publisher}{Brill},
  \bibinfo{address}{Leiden, The Netherlands}),
  \dourl{https://brill.com/view/title/24210}.

\bibitem[{Nagae \emph{et~al.}(2014)Nagae, Irino, Nisimura, Kawahara, and
  Patterson}]{nagae2014hearing}
\bibinfo{author}{Nagae, M.}, \bibinfo{author}{Irino, T.},
  \bibinfo{author}{Nisimura, R.}, \bibinfo{author}{Kawahara, H.},  and
  \bibinfo{author}{Patterson, R.~D.} (\textbf{\bibinfo{year}{2014}}).
  \enquote{\bibinfo{title}{Hearing impairment simulator based on compressive
  gammachirp filter}} in \emph{\bibinfo{booktitle}{Signal and Information
  Processing Association Annual Summit and Conference (APSIPA), 2014
  Asia-Pacific}}, \bibinfo{organization}{IEEE}, pp. \bibinfo{pages}{1--4},
  \dodoi{10.1109/APSIPA.2014.7041579}.

\bibitem[{Nejime and Moore(1997)}]{nejime1997simulation}
\bibinfo{author}{Nejime, Y.},  and \bibinfo{author}{Moore, B.~C.}
  (\textbf{\bibinfo{year}{1997}}). \enquote{\bibinfo{title}{Simulation of the
  effect of threshold elevation and loudness recruitment combined with reduced
  frequency selectivity on the intelligibility of speech in noise}}
  \bibinfo{journal}{The Journal of the Acoustical Society of America}
  \textbf{102}(1), \bibinfo{pages}{603--615}, \dodoi{10.1121/1.419733}.

\bibitem[{Nelson \emph{et~al.}(2001)Nelson, Schroder, and
  Wojtczak}]{nelson2001new}
\bibinfo{author}{Nelson, D.~A.}, \bibinfo{author}{Schroder, A.~C.},  and
  \bibinfo{author}{Wojtczak, M.} (\textbf{\bibinfo{year}{2001}}).
  \enquote{\bibinfo{title}{A new procedure for measuring peripheral compression
  in normal-hearing and hearing-impaired listeners}} \bibinfo{journal}{The
  Journal of the Acoustical Society of America} \textbf{110}(4),
  \bibinfo{pages}{2045--2064}, \dodoi{10.1121/1.1404439}.

\bibitem[{Oxenham and Plack(1997)}]{oxenham1997behavioral}
\bibinfo{author}{Oxenham, A.~J.},  and \bibinfo{author}{Plack, C.~J.}
  (\textbf{\bibinfo{year}{1997}}). \enquote{\bibinfo{title}{A behavioral
  measure of basilar-membrane nonlinearity in listeners with normal and
  impaired hearing}} \bibinfo{journal}{The Journal of the Acoustical Society of
  America} \textbf{101}(6), \bibinfo{pages}{3666--3675},
  \dodoi{10.1121/1.418327}.

\bibitem[{Patterson(1976)}]{patterson1976auditory}
\bibinfo{author}{Patterson, R.~D.} (\textbf{\bibinfo{year}{1976}}).
  \enquote{\bibinfo{title}{Auditory filter shapes derived with noise stimuli}}
  \bibinfo{journal}{The Journal of the Acoustical Society of America}
  \textbf{59}(3), \bibinfo{pages}{640--654}, \dodoi{10.1121/1.380914}.

\bibitem[{Patterson \emph{et~al.}(1982)Patterson, Nimmo-Smith, Weber, and
  Milroy}]{patterson1982deterioration}
\bibinfo{author}{Patterson, R.~D.}, \bibinfo{author}{Nimmo-Smith, I.},
  \bibinfo{author}{Weber, D.~L.},  and \bibinfo{author}{Milroy, R.}
  (\textbf{\bibinfo{year}{1982}}). \enquote{\bibinfo{title}{The deterioration
  of hearing with age: Frequency selectivity, the critical ratio, the
  audiogram, and speech threshold}} \bibinfo{journal}{The Journal of the
  Acoustical Society of America} \textbf{72}(6), \bibinfo{pages}{1788--1803},
  \dodoi{10.1121/1.388652}.

\bibitem[{Patterson \emph{et~al.}(2003)Patterson, Unoki, and
  Irino}]{patterson2003extending}
\bibinfo{author}{Patterson, R.~D.}, \bibinfo{author}{Unoki, M.},  and
  \bibinfo{author}{Irino, T.} (\textbf{\bibinfo{year}{2003}}).
  \enquote{\bibinfo{title}{Extending the domain of center frequencies for the
  compressive gammachirp auditory filter}} \bibinfo{journal}{The Journal of the
  Acoustical Society of America} \textbf{114}(3), \bibinfo{pages}{1529--1542},
  \dodoi{10.1121/1.1600720}.

\bibitem[{Sergeyenko \emph{et~al.}(2013)Sergeyenko, Lall, Liberman, and
  Kujawa}]{sergeyenko2013age}
\bibinfo{author}{Sergeyenko, Y.}, \bibinfo{author}{Lall, K.},
  \bibinfo{author}{Liberman, M.~C.},  and \bibinfo{author}{Kujawa, S.~G.}
  (\textbf{\bibinfo{year}{2013}}). \enquote{\bibinfo{title}{Age-related
  cochlear synaptopathy: an early-onset contributor to auditory functional
  decline}} \bibinfo{journal}{Journal of Neuroscience} \textbf{33}(34),
  \bibinfo{pages}{13686--13694}.

\bibitem[{Stone and Moore(1999)}]{stone1999tolerable}
\bibinfo{author}{Stone, M.~A.},  and \bibinfo{author}{Moore, B.~C.}
  (\textbf{\bibinfo{year}{1999}}). \enquote{\bibinfo{title}{Tolerable hearing
  aid delays. i. estimation of limits imposed by the auditory path alone using
  simulated hearing losses}} \bibinfo{journal}{Ear and Hearing} \textbf{20}(3),
  \bibinfo{pages}{182--192},
  \dourl{https://journals.lww.com/ear-hearing/Abstract/1999/06000/Tolerable_Hearing_Aid_Delays__I__Estimation_of.2.aspx}.

\bibitem[{Tsuiki \emph{et~al.}(2002)Tsuiki, Sasamori, Minami, Ichinohe, Murai,
  Murai, and Kawashima}]{tsuiki2002nihon}
\bibinfo{author}{Tsuiki, T.}, \bibinfo{author}{Sasamori, S.},
  \bibinfo{author}{Minami, Y.}, \bibinfo{author}{Ichinohe, T.},
  \bibinfo{author}{Murai, K.}, \bibinfo{author}{Murai, S.},  and
  \bibinfo{author}{Kawashima, H.} (\textbf{\bibinfo{year}{2002}}).
  \enquote{\bibinfo{title}{Age effect on hearing: a study on japanese}}
  \bibinfo{journal}{Audiology Japan (in Japanese)} \textbf{45}(3),
  \bibinfo{pages}{241--250}, \dodoi{10.4295/audiology.45.241}.

\bibitem[{Unoki \emph{et~al.}(2006)Unoki, Irino, Glasberg, Moore, and
  Patterson}]{unoki2006comparison}
\bibinfo{author}{Unoki, M.}, \bibinfo{author}{Irino, T.},
  \bibinfo{author}{Glasberg, B.}, \bibinfo{author}{Moore, B.~C.},  and
  \bibinfo{author}{Patterson, R.~D.} (\textbf{\bibinfo{year}{2006}}).
  \enquote{\bibinfo{title}{Comparison of the roex and gammachirp filters as
  representations of the auditory filter}} \bibinfo{journal}{The Journal of the
  Acoustical Society of America} \textbf{120}(3), \bibinfo{pages}{1474--1492}.

\bibitem[{Villchur(1974)}]{villchur1974simulation}
\bibinfo{author}{Villchur, E.} (\textbf{\bibinfo{year}{1974}}).
  \enquote{\bibinfo{title}{Simulation of the effect of recruitment on loudness
  relationships in speech}} \bibinfo{journal}{The Journal of the Acoustical
  Society of America} \textbf{56}(5), \bibinfo{pages}{1601--1611},
  \dodoi{10.1121/1.1903484}.

\bibitem[{Zurek and Desloge(2007)}]{zurek2007hearing}
\bibinfo{author}{Zurek, P.~M.},  and \bibinfo{author}{Desloge, J.~G.}
  (\textbf{\bibinfo{year}{2007}}). \enquote{\bibinfo{title}{Hearing loss and
  prosthesis simulation in audiology}} \bibinfo{journal}{The Hearing Journal}
  \textbf{60}(7), \bibinfo{pages}{32--38},
  \dodoi{10.1097/01.HJ.0000281789.77088.b6}.

\end{thebibliography}

\end{document}